# A New Exploration into Chinese Characters: from Simplification to Deeper Understanding

Wen G. Gong*

## 1 Abstract

This paper presents a novel approach to Chinese characters through the lens of physics, network analysis, and natural systems. Computational analysis of over 6,000 characters identified 422 elemental characters (元字) as fundamental building blocks. Using a physics-inspired “Zi-Matrix” model, we analyzed character structure across eleven spatial positions, revealing systematic patterns in component relationships and semantic extension.

Our research demonstrates that Chinese characters exhibit properties of natural systems: emergent complexity, self-organization, and adaptive resilience. The Fibonacci sequence provides an organizing framework for understanding character evolution, from simple pictographs to sophisticated abstractions. Case studies of character families and semantic networks show how meaning radiates from concrete to abstract domains while maintaining coherent principles.

By viewing Chinese characters as a living system, this research transcends mere simplification to reveal how human cognition organizes and transmits knowledge. While the elemental character set reduces memorization burden, it also illuminates profound connections between language, thought, and natural patterns. Chinese characters emerge not just as tools for communication, but as windows into human understanding. This perspective, combined with AI-assisted learning approaches, promises to transform language education from knowledge mastery to meaning discovery, bridging traditional wisdom with modern computational methods.

## 2 Motivation

Learning 汉字 (Hànzì) and, by extension, the Chinese language presents a unique and substantial challenge for learners, particularly those whose native languages utilize alphabetic systems. The difficulty stems primarily from the logographic nature of 汉字, where each character represents a morpheme (or word) rather than a sound. Unlike phonetic scripts, there’s often no readily apparent connection between a character’s visual form and its pronunciation, demanding rote memorization of thousands of distinct characters to achieve basic literacy. This is compounded by the tonal nature of Mandarin Chinese, where changes in pitch can drastically alter the meaning of a word, adding another layer of complexity. Furthermore, the sheer volume of characters, the subtle nuances in stroke order and character composition, and the existence of both traditional and simplified forms, require a sustained and dedicated learning effort over a prolonged

*Corresponding author: wen_gong@vanguard.com

period, making the acquisition of fluency in written and spoken Chinese a significantly time-intensive undertaking compared to many other languages.

Inspired by the success of reductionism in science, such as the discovery of fundamental particles in physics and the organization of the periodic table in chemistry, this research seeks to apply similar principles to the learning of 汉字. This involves three key approaches:

- **Network Analysis**: Leveraging computer science and network analysis techniques, this study aims to uncover the hidden relationships and underlying structures within the large collection of Chinese characters. By mapping these connections, the research hopes to reveal patterns and simplify the seemingly chaotic complexity of the character system.
- **Artificial Intelligence (AI) Assistance**: Recognizing the burden of rote memorization in traditional 汉字 learning, this research explores the use of AI to alleviate this challenge. The goal is to develop AI-powered tools that can assist learners in memorizing the form, pronunciation, and meaning of characters, as well as their complex interactions and contextual usage.
- **Improve learning and learning expeirence**: The overarching research goal, built upon by applying network analysis and AI asisstance, is to reduce the learning burdens for students and enrich their learning experiences.

By combining these computational and AI-driven approaches with a reductionist perspective, the research aims to provide a novel exploration into understanding and learning 汉字, ultimately making the process more efficient, intuitive, and enjoyable for learners.

# 3 Computational Network Analysis on Chinese Characters

The foundational work Shuowen Jiezi (《说文解字》), compiled by Xu Shen (许慎) during the Eastern Han Dynasty, represents the first systematic attempt to analyze the structure and etymology of Chinese characters [1, 2, 3]. Xu Shen identified 540 radicals (部首, bùshǒu) and used his "six scripts" (六书, liùshū) theory to explain character formation, categorizing characters based on their composition: pictographs (象形), ideographs (指事), compound ideographs (会意), phono-semantic compounds (形声), and two less-common categories (转注 and 假借). While groundbreaking, the Shuowen Jiezi's analysis, focused on the Small Seal Script (小篆), doesn't perfectly reflect the modern forms of many characters, and its 540 radicals, while valuable, do not always represent the smallest, irreducible components.

The later Zihui dictionary and the Kangxi Dictionary (康熙字典) refined and reduced this system, ultimately settling on the 214 Kangxi radicals [4], which serve as a standard indexing system. These radicals, based primarily on shared visual components, often relating to meaning (semantic radicals, or 形旁, xíngpáng), provide a method for dictionary lookups. While many radicals also hint at pronunciation (phonetic radicals, or 声旁, shēngpáng), this is not always reliable. The Kangxi system, though widely used, has limitations: ambiguous categorization, inconsistent semantic and phonetic roles, and some radicals that, due to their complexity or low frequency of occurrence, are not always the most informative units for understanding character structure.

Building upon the legacy of Xu Shen and Kangxi systems, the author has developed a web application called "ZiNets". With this tool and computational network structure, we decomposed 6190 chinese characters [5] (including 3910 HSK common characters [6]). Elemental characters were identified as those components appearing with a frequency above a defined threshold.

To systematically analyze the structure of Chinese characters, we introduce a novel spatial decomposition model, the "Zi-Matrix." This model represents each character as a matrix of up to eleven distinct positional components. These positions are defined as: Top (上), Bottom (下), Left (左), Right (右), Center (中), Top-Left (左上), Top-Right (右上), Bottom-Left (左下), Bottom-Right (右下), Center-Inside (中内), and Center-Outside (中外). Each position within the matrix can be either occupied by a specific character component (a radical, a stroke, or a more complex sub-component) or remain empty. This 11-component matrix allows for a consistent and structured representation of the spatial relationships between the constituent parts of any Chinese character, regardless of its complexity.

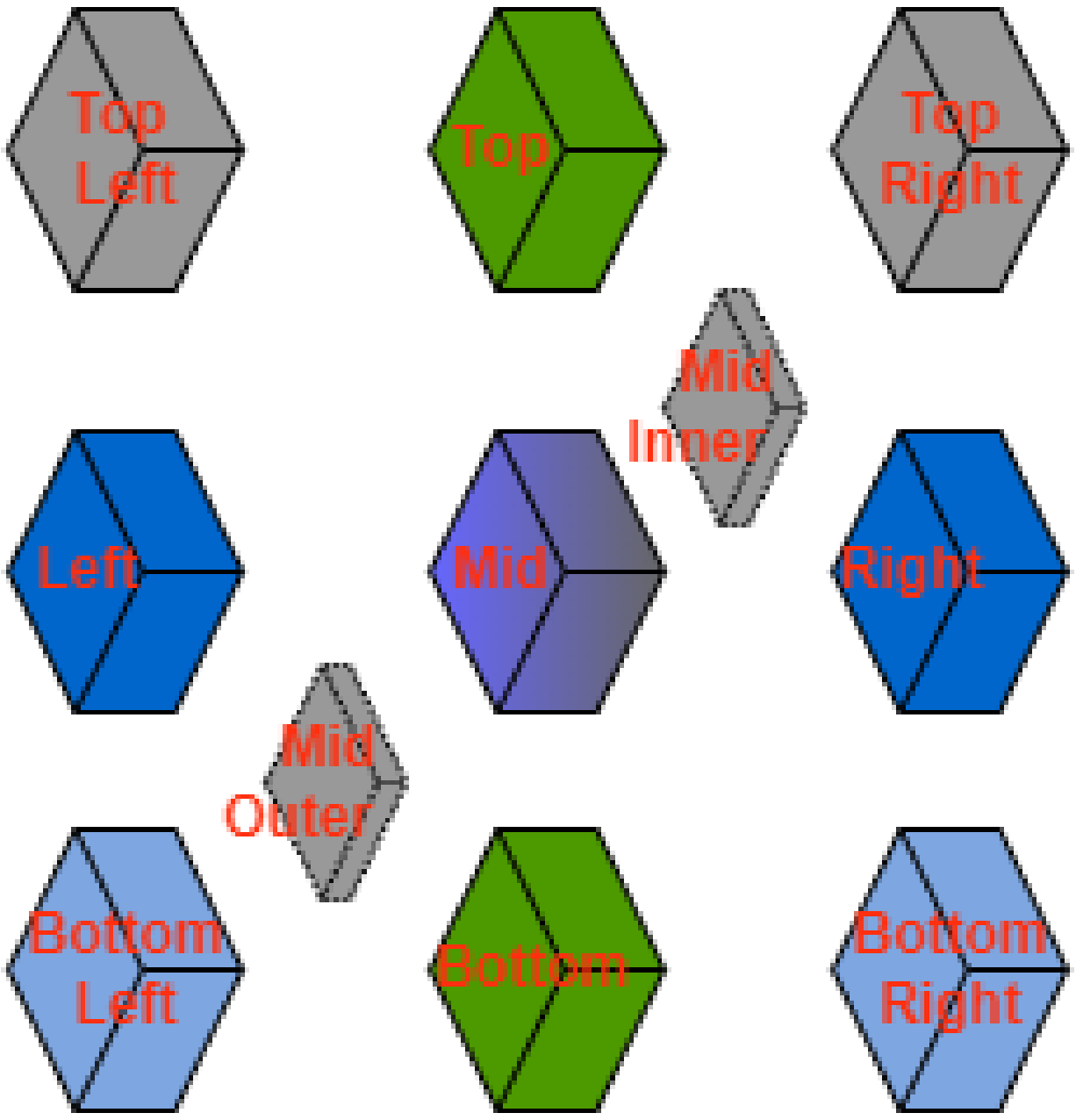

*Figure 1: Schematic representation of the Zi-Matrix spatial decomposition model for Chinese character network analysis.*

The decomposition of each character into the Zi-Matrix is performed manually using a hierarchical approach. This process involves a series of decomposition steps. First, the character is broken down into its major components, assigning each to its appropriate position within the matrix. If a component itself is complex, it is further decomposed, recursively, into its constituent parts, again assigning them to positions within a sub-matrix representing that component. This hierarchical decomposition continues until the most fundamental components – those that cannot be reasonably further divided – are identified, as illustrated by Figure 2. These fundamental components become candidates for the "elemental character" set. This manual, hierarchical approach ensures a consistent analysis of character structure, leveraging expert linguistic knowledge to guide the decomposition.

## 4 Elemental Characters Analysis

Through manual decomposition of 6190 Chinese characters using the Zi-Matrix method, we identified 422 unique, irreducible components as "elemental characters" (元字). These components represent the fundamental building blocks that emerged during our hierarchical decomposition process. Table 1 presents these elemental characters organized by stroke count, with a clear distinction between traditional Kangxi radicals and newly identified components.

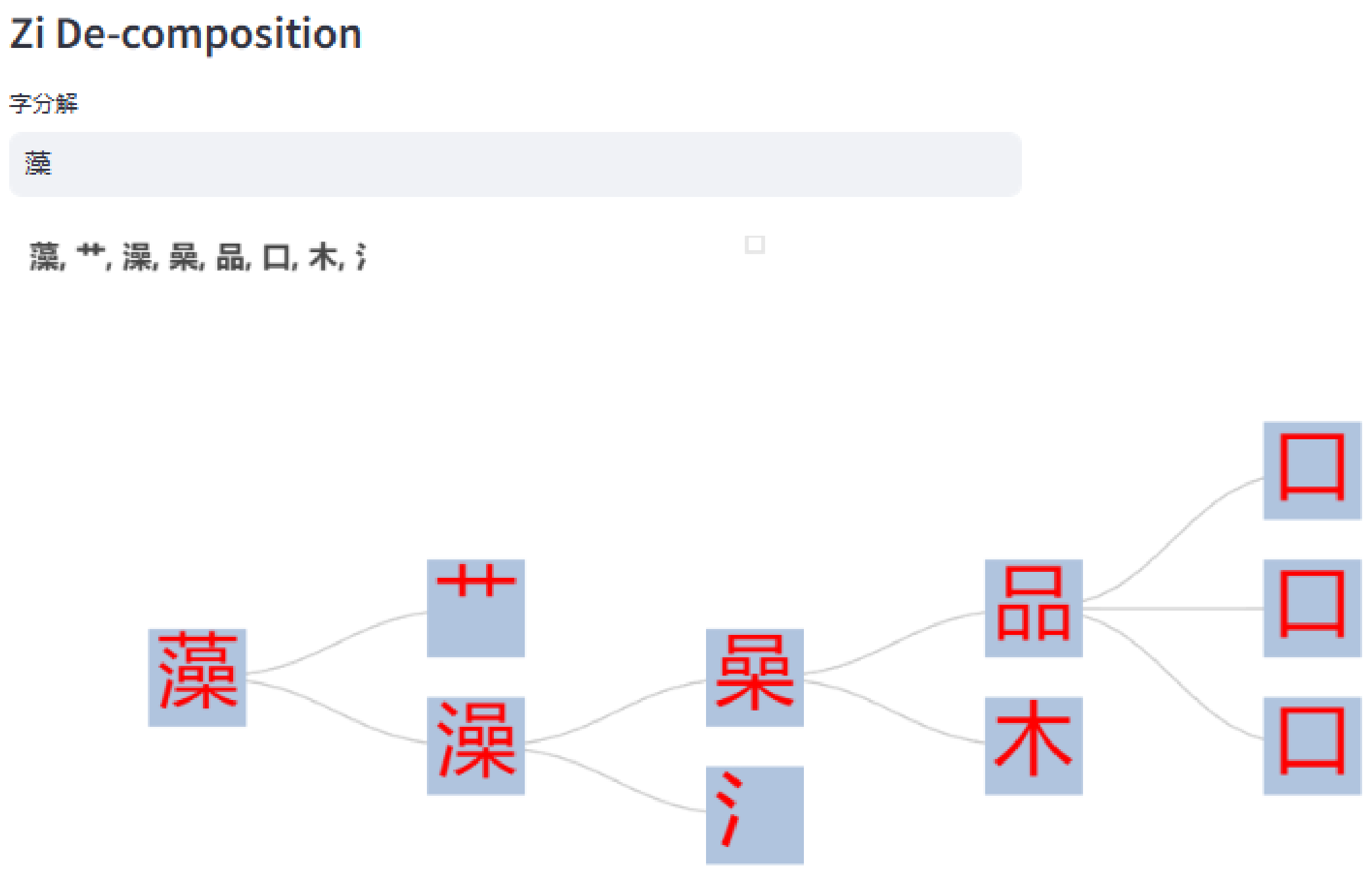

*Figure 2: The diagram of hierarchical decomposition tree illustrates the progressive breakdown of the character into its constituent components, reading from left to right. This visualization demonstrates the multi-level structural composition typical of complex Chinese characters.*

### 4.1 Elemental Characters (元字) by Stroke Count and Origin - Table 1

| **笔画数 (Stroke Count)** | **元字 (Kangxi Radicals)** | **元字 (New Radicals)** |
|---|---|---|
| 1 | 丨丶丿乀乁乚乛亅一乙 | |
| 2 | 丷亠亻冂冖冫凵⺈刂勹匚匸卩㔾厶讠二人儿入八几刀力匕十卜厂又 | ⺀乂龴𠂆丁七乃九了刁 |
| 3 | 夂夊宀尢屮巛廴彐彑彡彳忄扌氵丬纟艹辶阝饣口囗土士夕大女子寸小尸山川工己巾干幺广廾弋弓犭门飞马 | 亍兀⺌⺍阝万丈三上下与个丸义久么之乞也习于亏亡刃勺千叉及已乡才 |
| 4 | 攴攵殳灬爫爻爿牜⺩礻肉罒耂心戈戶户手支文斗斤方无日曰月木欠止歹毋比毛氏气水火爪父片牙牛犬王瓦见贝车长韦风 | 旡朩𠂉不丑专中丰为乌云五井亢今介仓以元公六内冈凶分办勾勿匀匹区升午友天太夫少尤尺屯巨巴开 |
| 5 | 氺疋疒癶罒衤钅母玄玉瓜甘生用田白皮皿目矛矢石示禾穴立鸟龙 | 乍刍戋𤴓且丘丙业东乎乐令兄兰冬出击包北半占卡去古句另只可台四央失头宁它尼市布平必斥旦未末本术正由甲申电 |
| 6 | ⺮𦍌聿艮虍覀竹米糸缶网羊羽老而耒耳肉臣自至臼舌舟色虫血行衣西页齐 | 囟尧䒑⺷交共各合吉向吕寺并庄式曲 |
| 7 | 豕豸酉卤舛角言谷豆赤走足身辛辰邑采里麦龟 | 㐬佥呙坙夅奂孚肙员杏甫良 |
| 8 | 隹黾金阜隶雨青非鱼齿 | 𠦝 幷其奉尚易東责 |
| 9 | 面革韭音食首香骨鬼 | 畐咸柬畏禺 |
| 10 | 高鬥 | |
| 11 | 麻鹿 | |
| 12 | 黍黄黑 | |
| 13 | 鼓 | |
| 14 | 鼻 | |

The identified elemental character set comprises 245 components derived from the traditional Kangxi radicals system and introduces 177 additional components. This expansion reflects our more granular approach to character decomposition, where some traditionally grouped radicals are treated as individual elements. These newly identified components, though not traditionally recognized as independent units in standard dictionaries, serve crucial phonetic and/or semantic roles in character formation. For instance, components like 禺 and 乍, while not classified as radicals in traditional systems, demonstrate consistent semantic contributions in compound characters, as we will explore in our case studies.

Our elemental character set reveals a finer granularity in Chinese character composition than the traditional Kangxi system. It encompasses three main types of components: familiar semantic radicals (e.g., 氵 [water], 木 [wood], 日 [sun], 月 [moon], 心 [heart], 手 [hand], 口 [mouth]), basic structural elements (e.g., 一, 丨, 丿, 丶, 乙, 口, 凵, 冂), and frequently occurring components that carry both phonetic and semantic information (e.g., 方, 占, 且, 戋, 乍, 禺, 尧, 佥). Some components inherited from the Kangxi system, such as 鼓 (drum), while historically significant, may be reconsidered for practical modern applications. This comprehensive set of elemental characters provides a more nuanced foundation for understanding and teaching Chinese character composition, potentially simplifying the learning process while maintaining semantic integrity.

### 4.2 Elemental Characters: Occurrence Frequency and Categorization

Analysis of high-frequency elemental characters (with occurrences above 23, a threshold chosen to include 气 (qi meaning air, breath and energy), given its fundamental importance in Chinese philosophy and culture) reveals clear patterns in character composition. The frequency distribution shows strong clustering around fundamental human concepts and natural elements. For instance, in human-related categories (人-), elements like 口 (mouth, 300 occurrences) and 手 (hand, 261 occurrences) show remarkably high usage rates, reflecting their importance in expressing human actions and experiences. Similarly, in the natural world categories, 水 (water, 377 occurrences) and 木 (wood, 324 occurrences) demonstrate high frequencies, indicating their crucial role in character formation related to natural phenomena. Notably, 亻 (human radical, 213 occurrences) and 女 (female, 137 occurrences) also show high frequencies, underlining the human-centric nature of character formation.

The categorization system reveals a hierarchical organization centered on major conceptual domains. The system distinguishes between human-centered categories (人-系列) including physiological (生理), psychological (心理), and behavioral (行) aspects; natural elements (天文-) including the traditional five elements (金木水火土); and categories for flora (植物-), fauna (动物-), mathematical concepts (数理-), and abstract concepts (概念-). This classification not only reflects traditional Chinese philosophical understanding of the world but also provides a systematic framework for understanding character composition. Notably, the frequency distribution within these categories suggests that characters related to human experience and basic natural elements form the core building blocks of the writing system, while more specialized or abstract concepts show lower frequencies. The inclusion of 气 (24 occurrences) in this analysis proves particularly significant, as it represents a threshold case that bridges fundamental philosophical concepts with practical character formation patterns.

| | A | B | C | D | E | F | G | H | I | J |
|---|---|---|---|---|---|---|---|---|---|---|
| 1 | zi | frequency | category | radical | strokes | zi | frequency | category | radical | strokes |
| 2 | 亻 | 213 | 人- | Y | 2 | 气 | 24 | 天文- | | 4 |
| 3 | 女 | 137 | 人- | | 3 | 日 | 125 | 天文-日 | | 4 |
| 4 | 人 | 46 | 人- | | 2 | 月 | 154 | 天文-月, 人-生理 | | 4 |
| 5 | 忄 | 116 | 人-心理 | Y | 3 | 钅 | 88 | 天文-金 | Y | 5 |
| 6 | 心 | 84 | 人-心理 | | 4 | 木 | 324 | 天文-木 | | 4 |
| 7 | 口 | 300 | 人-生理 | | 3 | 氵 | 377 | 天文-水 | Y | 3 |
| 8 | 言 | 36 | 人-生理 | | 7 | 冫 | 29 | 天文-水 | Y | 2 |
| 9 | 讠 | 102 | 人-生理 | Y | 2 | 火 | 85 | 天文-火 | | 4 |
| 10 | 目 | 79 | 人-生理 | | 5 | 土 | 143 | 天文-土 | | 3 |
| 11 | 扌 | 261 | 人-生理 | Y | 3 | 阝 | 153 | 地理-土 | Y | 3 |
| 12 | 又 | 73 | 人-生理 | | 2 | 石 | 71 | 地理-土 | | 5 |
| 13 | 攵 | 59 | 人-生理 | Y | 4 | 山 | 65 | 地理-土 | | 3 |
| 14 | 疒 | 74 | 人-生理 | Y | 5 | 艹 | 284 | 植物- | Y | 3 |
| 15 | 辶 | 90 | 人-行 | Y | 3 | 纟 | 111 | 植物- | Y | 3 |
| 16 | 足 | 76 | 人-行 | | 7 | ⺮ | 108 | 植物- | Y | 6 |
| 17 | 衤 | 52 | 人-衣 | Y | 5 | 糸 | 32 | 植物- | | 6 |
| 18 | 巾 | 55 | 人-衣 | | 3 | 虫 | 132 | 动物- | | 6 |
| 19 | 米 | 51 | 人-食 | | 6 | 鸟 | 62 | 动物- | | 5 |
| 20 | 禾 | 78 | 人-食 | | 5 | 隹 | 53 | 动物- | Y | 8 |
| 21 | 酉 | 46 | 人-食 | Y | 7 | 犭 | 51 | 动物- | | 3 |
| 22 | 宀 | 63 | 人-住 | Y | 3 | 马 | 44 | 动物- | | 3 |
| 23 | 王 | 113 | 社会- | | 4 | 羊 | 37 | 动物- | | 6 |
| 24 | 贝 | 59 | 社会- | | 4 | 一 | 73 | 数理-计算 | | 1 |
| 25 | 车 | 53 | 社会- | | 4 | 十 | 56 | 数理-计算 | | 2 |
| 26 | 田 | 51 | 社会- | | 5 | 八 | 28 | 数理-计算 | | 2 |
| 27 | 力 | 53 | 概念- | | 2 | 方 | 27 | 数理-时空 | | 4 |

*Figure 3: Frequency distribution and categorization of top elemental characters with occurrences above 23.*

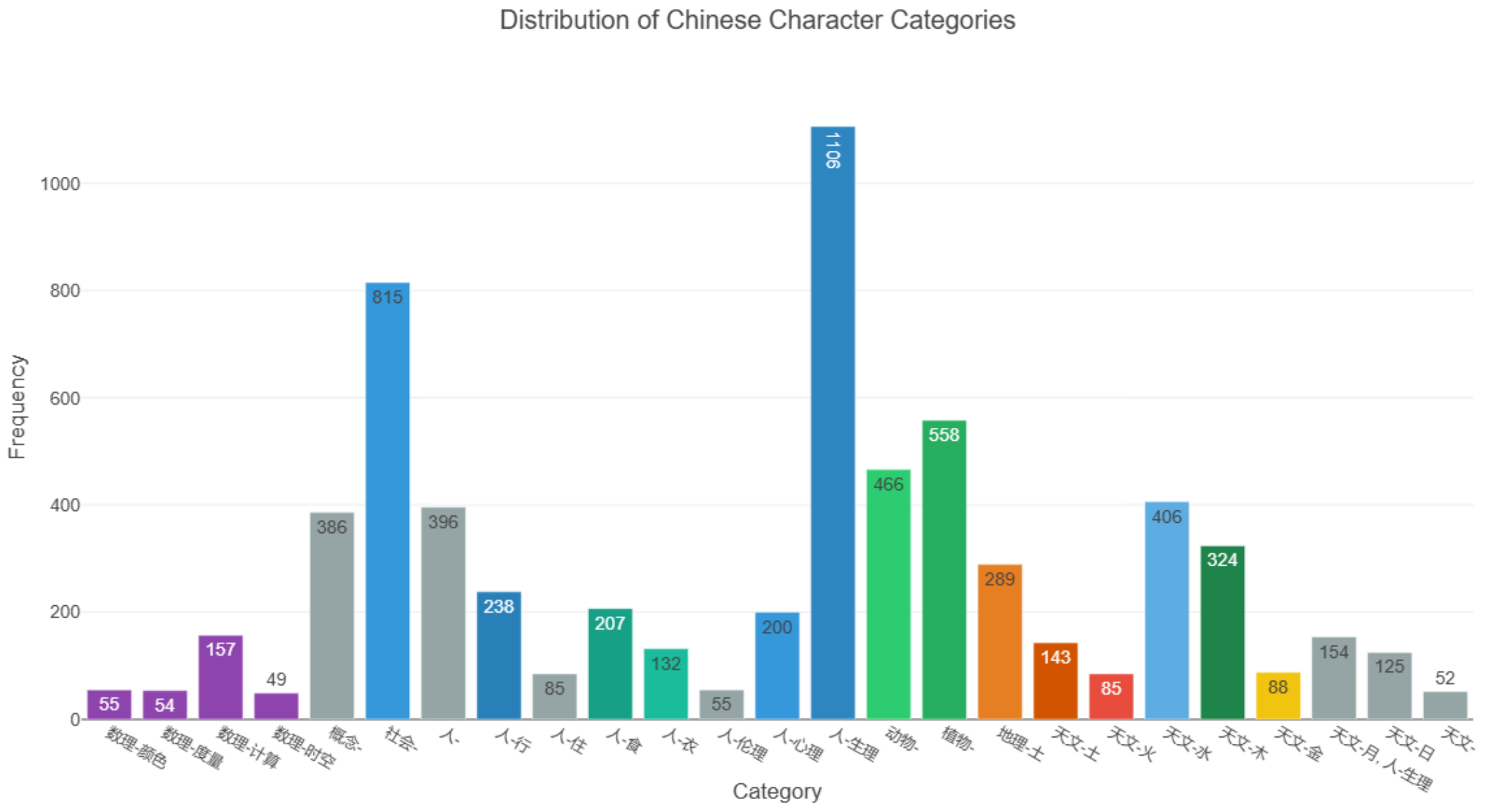

*Figure 4: Distribution of elemental characters across conceptual categories, showing frequency of occurrence ordered from cosmic to human to abstract realms.*

## 4.3 Visualizing Categorization

The visualization reveals a remarkable pattern in the distribution of Chinese characters that reflects ancient Chinese cosmological understanding. Beginning with celestial elements (天文-), the categories flow through natural phenomena to human experience, creating a narrative that resonates deeply with traditional Chinese philosophical principles. The striking prominence of human-related physiological characters (人-生理, 1106 occurrences) at the center of this distribution, bridged between natural elements and social constructs, echoes the classical Chinese view of humans as the connecting point between Heaven and Earth (天人合一). This central position is flanked by substantial representations in both natural domains—flora (植物-, 558 occurrences) and fauna (动物-, 466 occurrences)—and social spheres (社会-, 815 occurrences).

Within the astronomical/natural elements category (天文-), water (水, 406 occurrences) and wood (木, 324 occurrences) show notably higher frequencies than fire (火), metal (金), and earth (土), suggesting their greater significance in character formation. The systematic progression from celestial phenomena through natural elements, human experience, and finally to abstract mathematical concepts (数理-) reveals an elegant hierarchical structure that mirrors traditional Chinese cosmological ordering. This distribution pattern suggests an underlying organizational principle in Chinese character evolution that reflects both human cognitive development and natural world observations, a relationship that will be further explored through its connection to the Fibonacci sequence.

## 4.4 Character Compositional Patterns

Analysis of character compositions reveals patterns remarkably similar to molecular organization in nature. Just as atoms combine through fundamental interactions, elemental characters (元字) interact spatially to form more complex characters. The most prevalent arrangements mirror basic two-body interactions, appearing as phonetic-semantic compounds in various orientations: left-middle (L-M, 3,033 characters), up-down (U-D, 727 characters), up-middle (U-M, 688 characters), and left-right (L-R, 534 characters).

| | Compositional Pattern | Frequency |
|---|---|---|
| **0** | L-M | 3033 |
| **1** | U-D | 727 |
| **2** | U-M | 688 |
| **3** | L-R | 534 |
| **4** | M-R | 301 |
| **5** | U-M-D | 139 |
| **6** | LD-M | 120 |
| **7** | M-Mi | 118 |
| **8** | LU-M | 106 |
| **9** | M-D | 70 |
| **10** | D-LU-RU | 45 |
| **11** | LU-LD-R | 28 |
| **12** | L-M-R | 26 |
| **13** | U-LD-RD | 21 |
| **14** | M-Mo | 21 |
| **15** | L-RU-RD | 18 |
| **16** | U-L-R | 8 |
| **17** | Mi-Mo | 6 |
| **18** | LU-RU-LD-RD | 6 |

*Figure 5: Distribution of elemental character compositional patterns, showing frequency of each pattern type.*

These relative positions, while described in modern directional terms, reflect more fundamental spatial relationships that transcend the historical evolution from vertical to horizontal writing systems.

More complex patterns demonstrate sophisticated geometric arrangements analogous to multi-body interactions in physics. Three-component interactions manifest in triangular configurations like down-left_up-right_up (D-LU-RU, 45 characters) and left_up-left_down-right (LU-LD-R, 28 characters), while four-component interactions appear in symmetrical enclosure patterns, though less frequently (6 characters). This three-dimensional spatial organization distinguishes Chinese characters from the linear sequence of alphabetic writing systems, reflecting instead the multi-dimensional interactions observed in natural systems. The clear preference for simpler two-body arrangements, followed by decreasing frequency of more complex geometric patterns, mirrors nature's tendency toward efficient, stable configurations. This distribution provides valuable insights for both understanding character formation principles and developing effective learning strategies based on fundamental spatial relationships.

# 5 Storytelling By Character

Chinese characters are not merely symbols for recording language; they are sophisticated vehicles for storytelling that encode observations, wisdom, and natural principles within their structure. Building upon our analysis of elemental characters and their compositional patterns, we now explore how these basic units combine to tell stories at multiple scales - from individual character families to complete poems. Just as molecules tell the story of matter's organization through their bonds and interactions, Chinese characters reveal deeper narratives through their semantic relationships and evolving combinations. This section examines how meaning emerges through increasingly complex arrangements: from character families that share common elements, through network-like compound formations, to the crystalline structures of classical poetry. Each level demonstrates how Chinese writing, like natural systems, achieves remarkable efficiency in encoding information while maintaining profound aesthetic and philosophical coherence.

## 5.1 Case Studies - Composite Characters

### 5.1.1 The 日 Family

The Chinese character "日" (rì), meaning "sun" or "day", serves as a powerful illustration of how Chinese characters efficiently encode natural observations and knowledge. The character's evolution from its oracle bone pictograph—depicting a circular sun with a central dot—to its modern form demonstrates remarkable preservation of core visual concepts while gaining calligraphic efficiency.

| **Formula** | **Meaning** | **Natural Insight** | **Pinyin** |
|---|---|---|---|
| 日 + 月 = 明 | bright | Sun and moon as primary celestial light sources | míng |
| 日 + 正 = 是 | to be/truth | Overhead sun casts no shadows, revealing true form | shì |
| 知 + 日 = 智 | wisdom | The sun radiates energy without depletion, demonstrating how true wisdom transcends accumulated knowledge to understand principles of sustainable benefit | zhì |
| 日 + 日 + 日 = 晶 | crystal/bright | Intensification of light/clarity through repetition | jīng |
| 门 + 日 = 间 | space/interval | Light revealing space between door frames | jiān |
| 日 + 寸 = 时 | time | Sun's shadow measure in ancient timekeeping | shí |
| 日 + 生 = 星 | star | Stars as celestial light producers | xīng |
| 丿 + 日 = 白 | white | Pure light emanating from sun | bái |
| 日 + 一 = 旦 | dawn | Sun rising above horizon | dàn |
| 九 + 日 = 旭 | rising sun | Multiple rays of morning sunlight | xù |
| 日 + 十 = 早 | early | Sun above treetops at dawn | zǎo |
| 日 + 干 = 旱 | drought | Intense sun causing dryness | hàn |
| 三 + 八 + 日 = 春 | spring | Sunlight causing seeds to break soil, implying the Spring season | chūn |

The character 春 (spring) exemplifies efficient semantic encoding through its components: 三 (representing multiple aspects - abundance of sprouting life, layers of frozen earth), 八 (break/split), and 日 (sun).

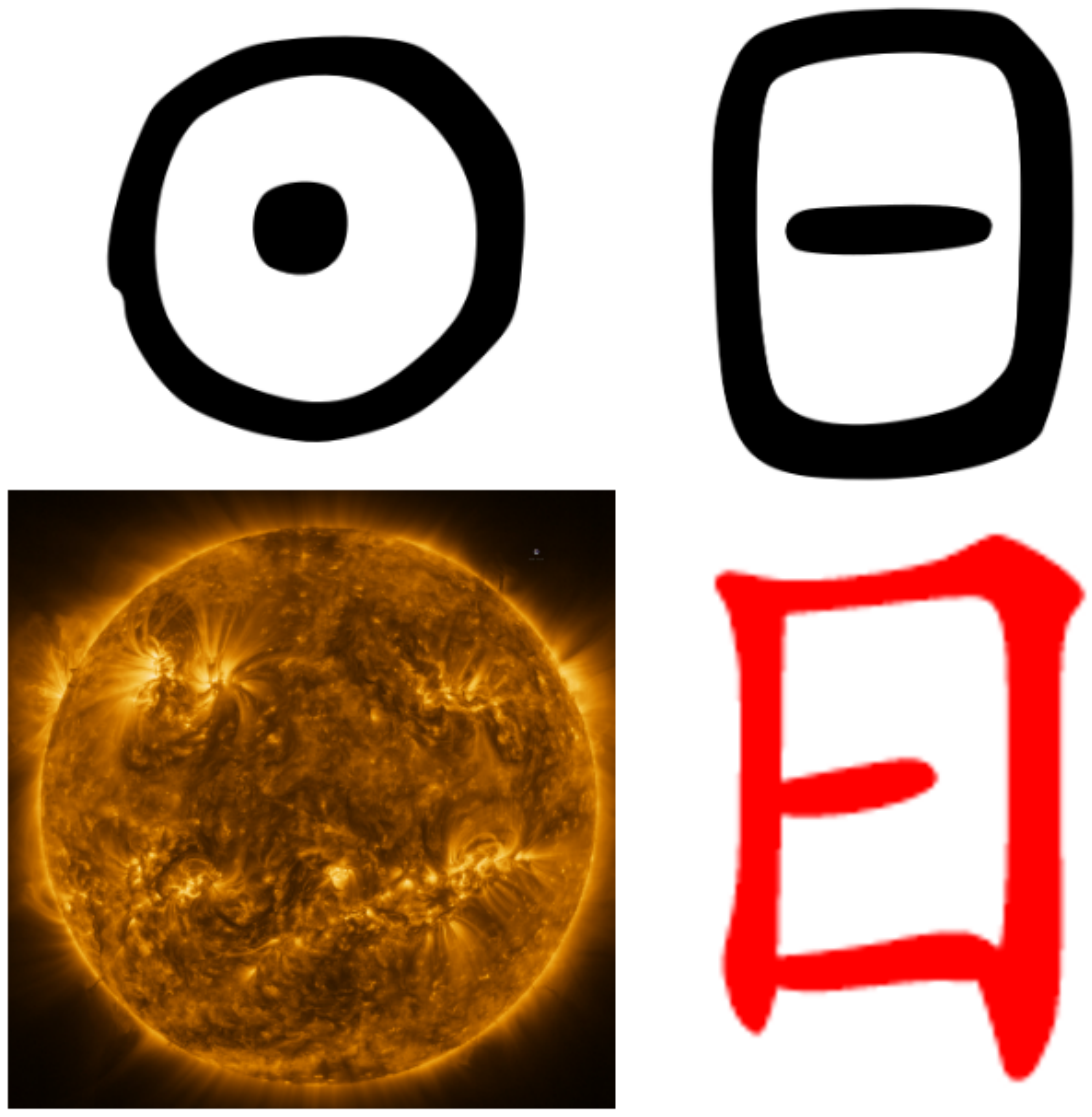

*Figure 6: Evolution of the character 日 (sun) from oracle bone pictograph to modern form, alongside actual solar image.*

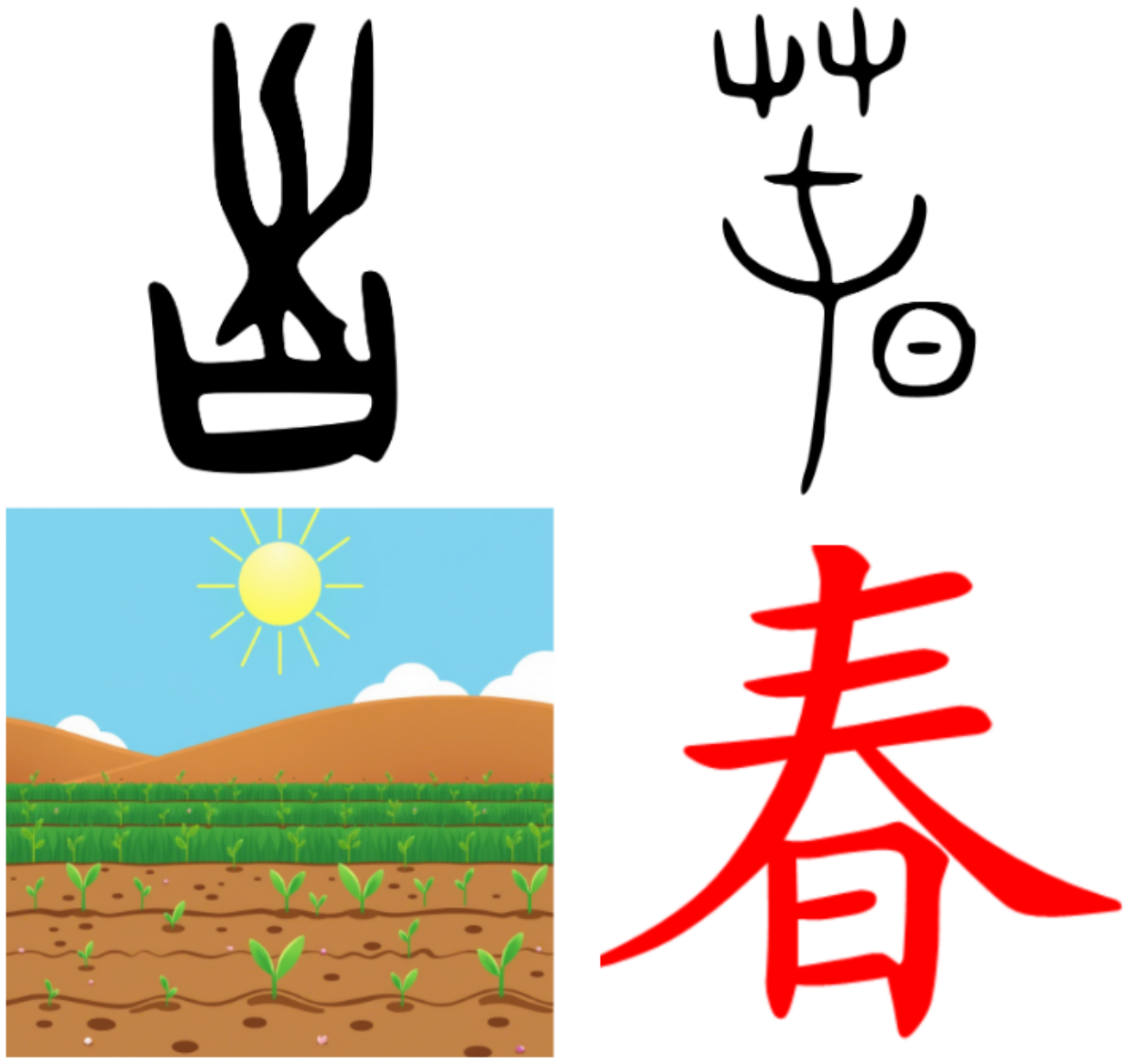

*Figure 7: The character 春 (spring) captures multiple aspects of seasonal transition.*

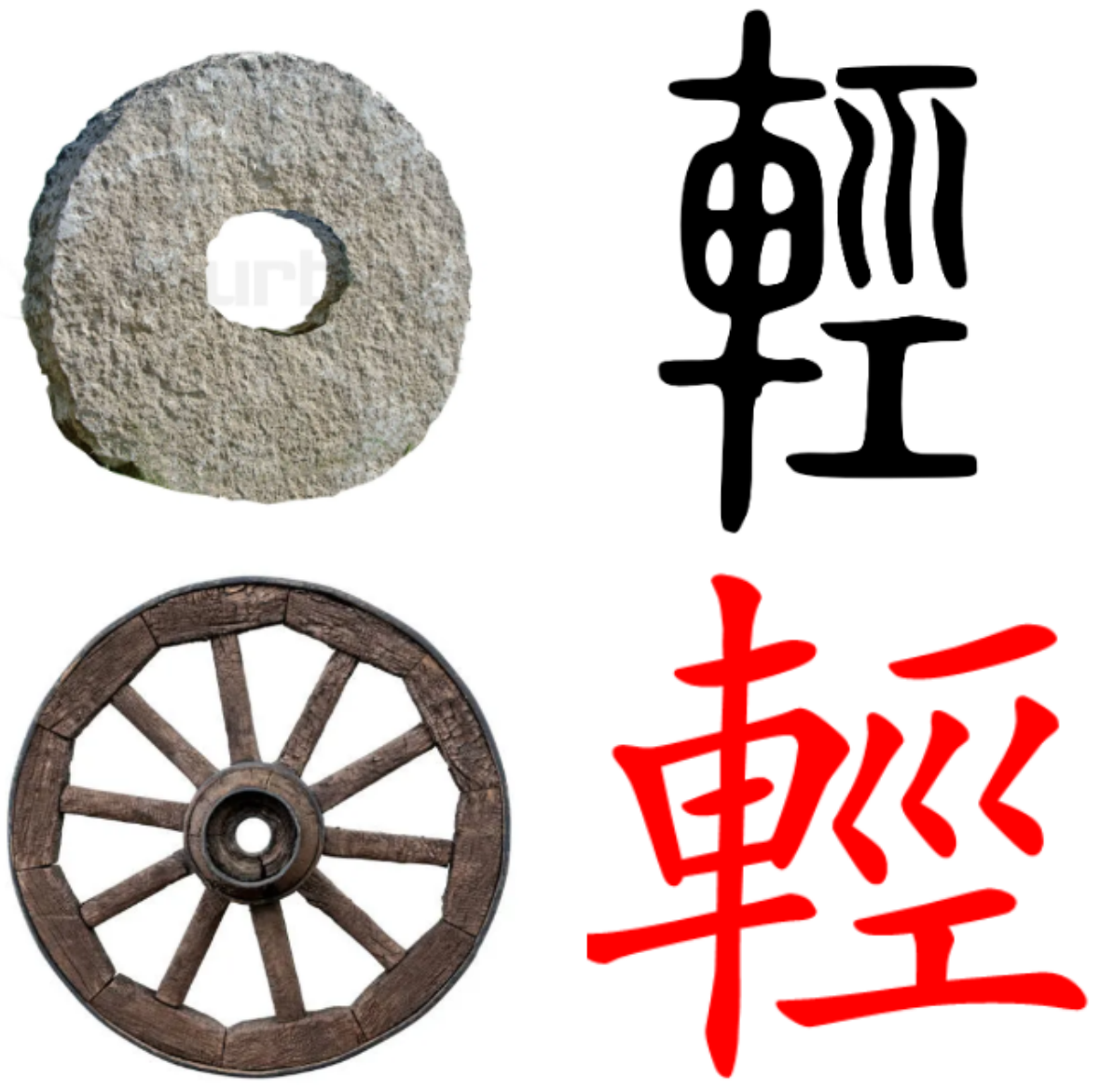

*Figure 8: The character 巠 encodes the evolution from material mass to structural efficiency.*

Together, these elements capture the dynamics of spring's arrival - the interaction between strengthening sunlight, earth's layers, and emerging life. This structure demonstrates how Chinese characters can layer multiple related meanings within a single, coherent form.

#### 5.1.2 The 巠 Family

The character 巠 exemplifies how Chinese writing preserves fundamental principles of human innovation. Beyond its basic meaning of "core" and "lightness," this character documents a critical transition in human understanding: the discovery that effectiveness comes from optimized structure rather than raw mass. The progression from solid stone wheels to spoked wooden designs represents both technological advancement and conceptual breakthrough - the recognition that core principles often emerge through reduction rather than addition.

| **Formula** | **Meaning** | **Natural Insight** | **Pinyin** |
|---|---|---|---|
| 气 + 巠 = 氢 | hydrogen | Simplest element with single proton core | qīng |
| 艹 + 巠 = 茎 | plant stem | Essential structural support maintaining plant integrity | jīng |
| 纟 + 巠 = 经 | classic texts/meridian | Core texts that weave through cultural/anatomical fabric | jīng |
| 巠 + 力 = 劲 | strength/vigor | How core fibers generate physical power and strength | jìn |
| 巠 + 页 = 颈 | neck | Crucial connecting structure between body and brain | jǐng |
| 彳 + 巠 = 径 | path/radius | Most direct route from center to periphery | jìng |
| 车 + 巠 = 轻 | light-weight | Structural efficiency through essential components | qīng |

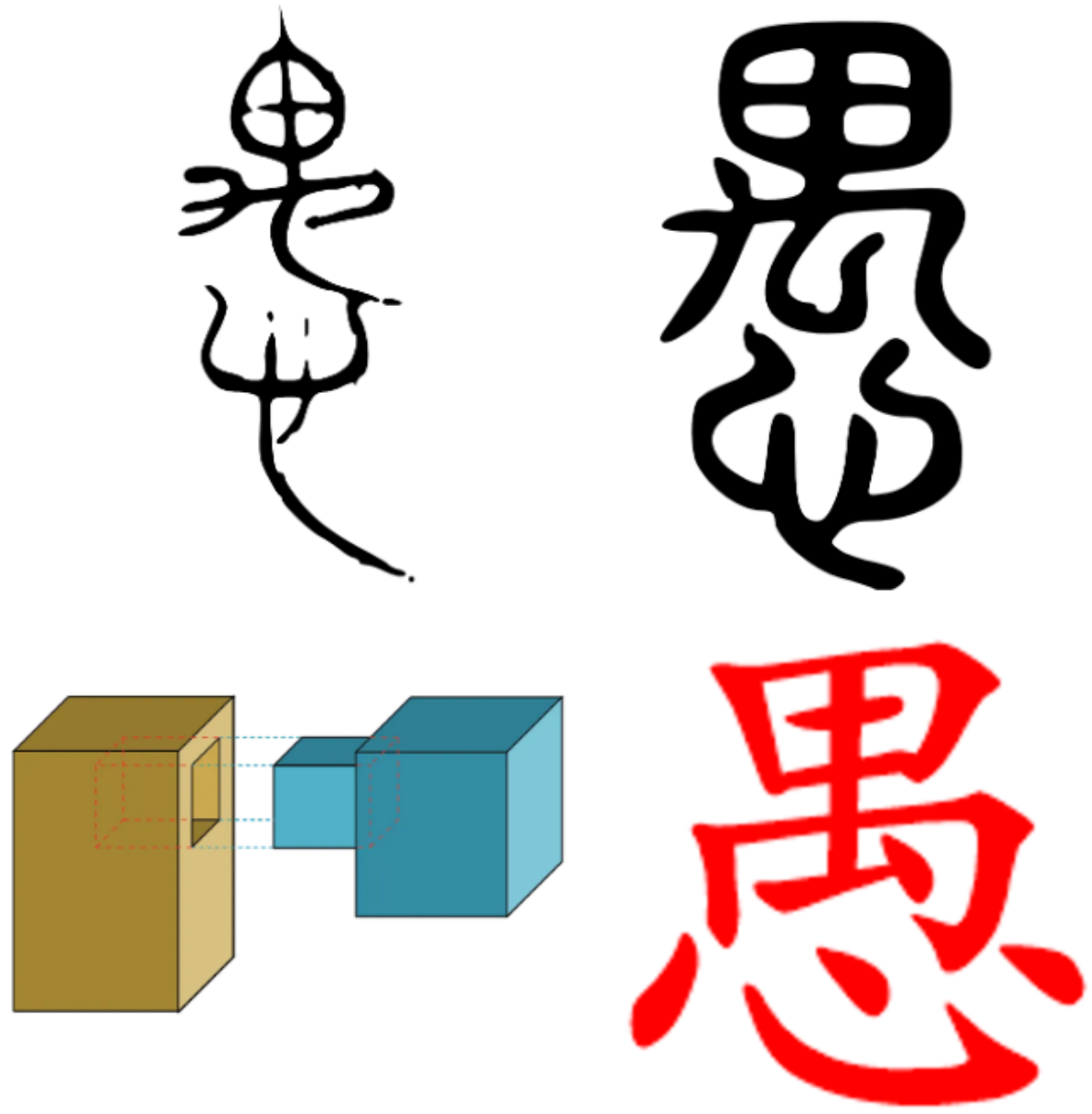

*Figure 9: The character 禺 evolution, depicting fundamental concept of joining/coupling.*

---

This character family demonstrates the Chinese writing system's capacity to encode and transmit complex principles across generations. Each derivative character extends the core concept of structural efficiency into different domains, from atomic physics (氫) to biological systems (莖), preserving not just the end results but the underlying principles of innovation.

### 5.1.3 The 禺 Family

The character 禺 represents a fundamental concept of joining or coupling in Chinese writing, functioning as a semantic force carrier for various types of connections and interactions.

| **Formula** | **Meaning** | **Natural Insight** | **Pinyin** |
|---|---|---|---|
| 亻 + 禺 = 偶 | 1. couple, partner; 2. chance, accident; 3. even number | Captures both intentional pairing and chance encounters, while "even numbers" suggests balanced states | ǒu |
| 宀 + 禺 = 寓 | dwelling, metaphor | Physical space enabling connections | yù |
| 辶 + 禺 = 遇 | encounter, meet | Dynamic interaction through movement | yù |
| 禺 + 心 = 愚 | inability to connect | Cognitive dimension of connection | yú |
| 阝 + 禺 = 隅 | corner, intersection | Spatial manifestation of joining | yú |
| 耒 + 禺 = 耦 | linked roots | Natural systems demonstrating connection | ǒu |

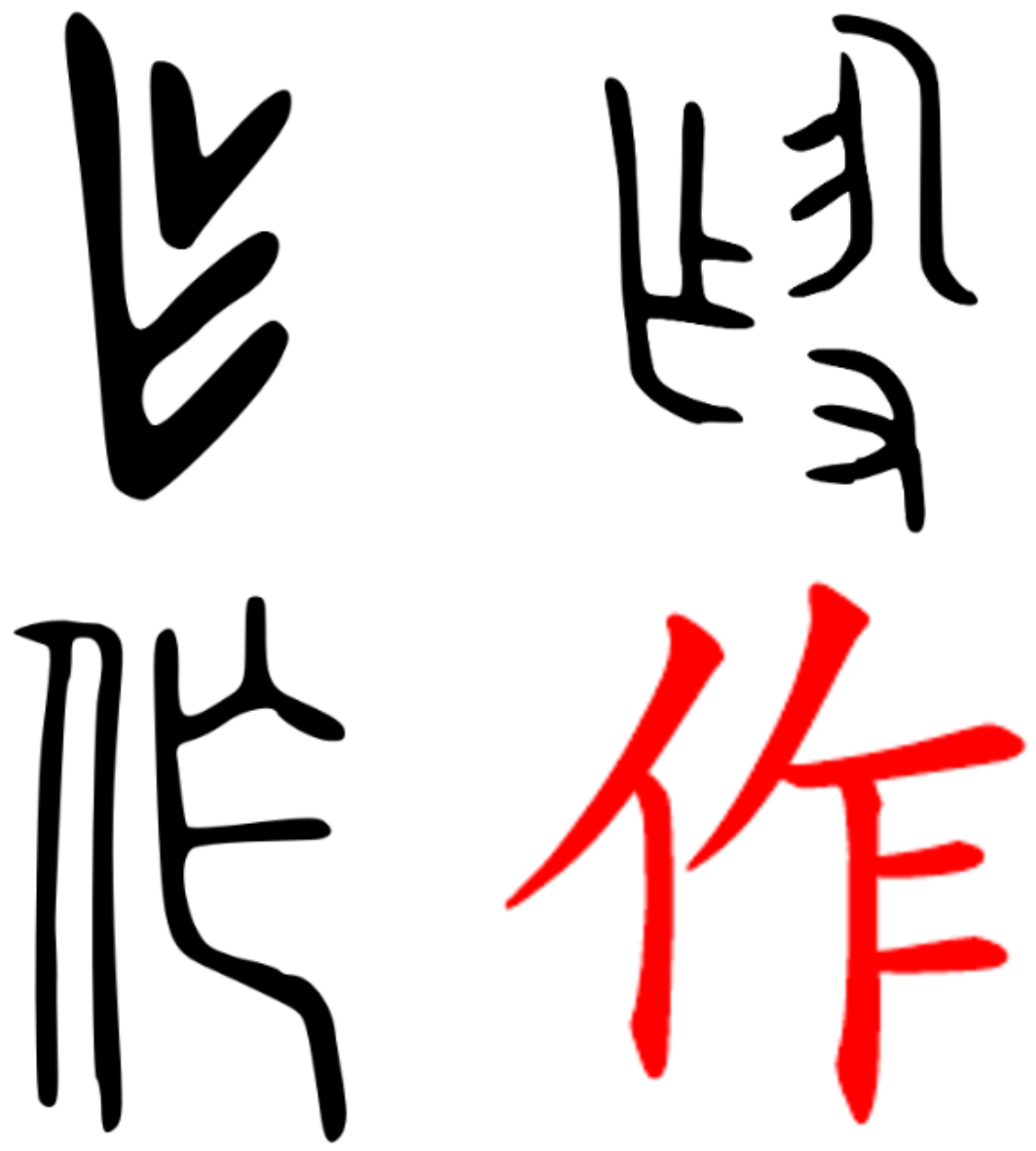

*Figure 10: Evolution of character 乍 (work/make) from oracle bone to modern form.*

---

This character family demonstrates how Chinese writing conceptualizes different types of connections - from physical joining to spatial relationships to cognitive associations - all derived from a single fundamental concept of coupling.

#### 5.1.4 The 乍 Family

The character 乍 represents fundamental concepts of work, creation, and transformation. Its historical forms suggest the process of making or transforming materials, functioning as a semantic force carrier for various types of productive activity.

| Formula | Meaning | Natural Insight | Pinyin |
|---|---|---|---|
| 亻 + 乍 = 作 | to make, to do | Direct human productive activity | zuò |
| 日 + 乍 = 昨 | yesterday | Time transformed through work completed | zuó |
| 乍 + 心 = 怎 | how? | Mental process of problem-solving | zěn |
| 火 + 乍 = 炸 | to explode, to fry | Energy transformation in different scales and shapes | zhà |
| 讠 + 乍 = 诈 | to deceive | Misuse of effort in communication | zhà |
| 口 + 乍 = 咋 | how | Verbal expression of questioning process | zǎ |
| 酉 + 乍 = 酢 | vinegar | Chemical transformation through work | zuò |
| ⺮ + 乍 = 笮 | to press | Physical work applied to materials | zé |
| 穴 + 乍 = 窄 | to hallow, to narrow | Spatial transformation through constraint | zhǎi |

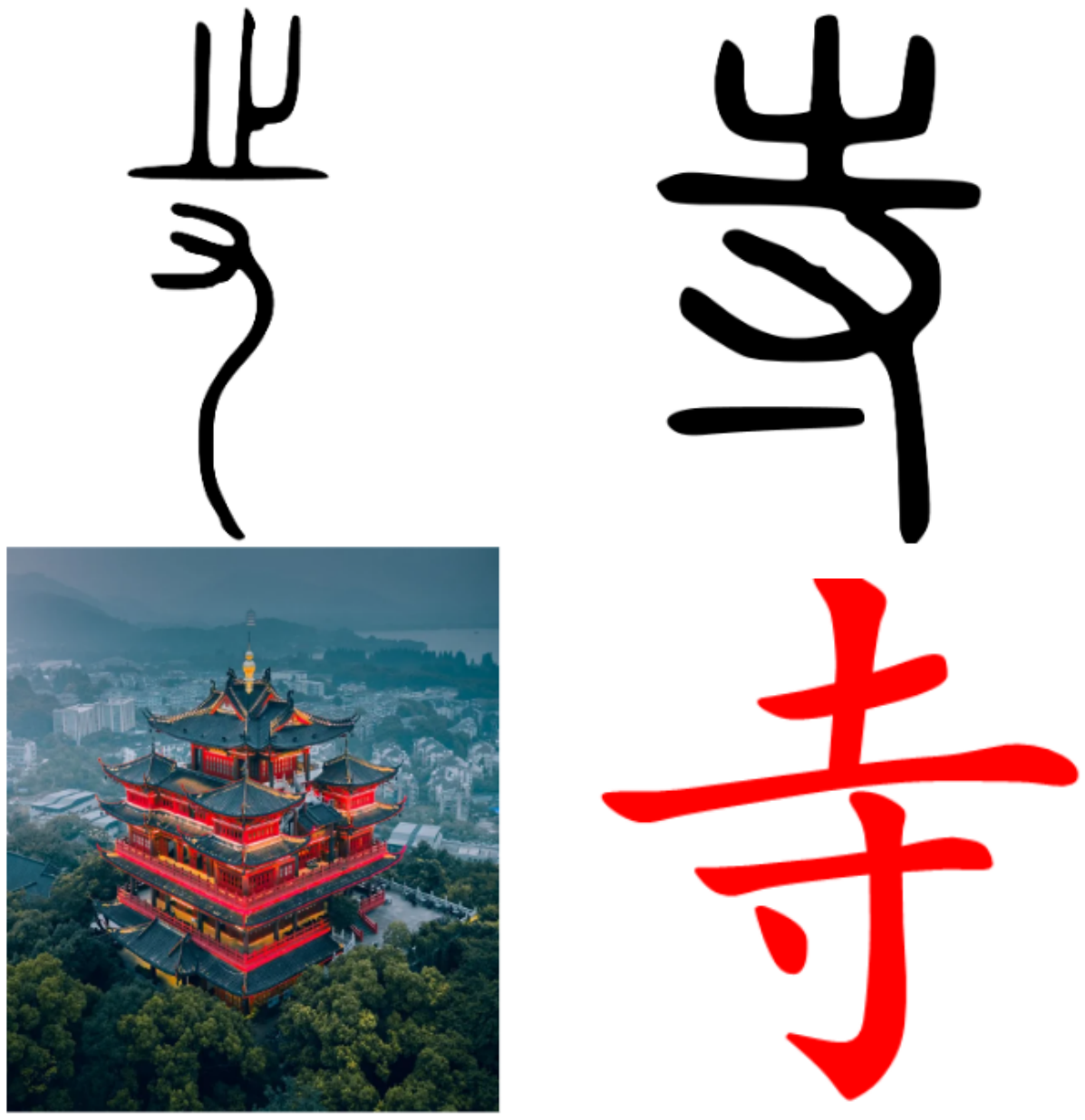

*Figure 11: Evolution of character 寺 (temple) from oracle bone script to modern form, with example of traditional temple architecture.*

---

This family demonstrates how a basic concept of work/transformation generates characters spanning physical, chemical, mental, temporal, and social domains - much like how energy manifests in different forms throughout nature.

#### 5.1.5 The 寺 Family

The character 寺, originally depicting a temple or place of authority, functions as a semantic carrier representing ordered structure and measured interaction. Its architectural origins in temple design suggest principles of balance, hierarchy, and measured sacred space.

| **Formula** | **Meaning** | **Natural Insight** | **Pinyin** |
|---|---|---|---|
| 日 + 寺 = 時 | time as cosmic law | Temple's ritualistic measurement of sun's movement reveals time as an inescapable cosmic order that governs all existence - a fundamental principle that commands universal respect and submission | shí |
| 牛 + 寺 = 特 | special / extraordinary | Cattle as precious agricultural asset offered to temple represents highest form of ritual sacrifice and distinction | tè |
| 亻 + 寺 = 侍 | to serve with ritual propriety | Models service relationships on temple's hierarchical spatial organization of inner/outer courts | shì |
| 扌 + 寺 = 持 | to uphold / maintain with stability | Reflects temple architecture's balanced distribution of structural forces and sustained equilibrium | chí |
| 彳 + 寺 = 待 | to await at proper position | Like prescribed positions in temple processions, represents disciplined positioning in space-time | dài |
| ⺮ + 寺 = 等 | ordered ranking / classification | Mirrors temple architecture's clear demarcation of sacred spaces by rank and function | děng |
| 山 + 寺 = 峙 | to tower with dignity | Embodies temple pagoda's vertical presence as symbol of spiritual and moral elevation | zhì |
| 忄 + 寺 = 恃 | to depend upon with confidence | Like temple foundations, represents reliable structural support extended to mental realm | shì |
| 讠 + 寺 = 诗 | poetry as sacred architecture | Words arranged with temple-like precision to create spaces of transcendent meaning | shī |

The 寺 family reveals how temple architecture's principles of structure and order extend into cosmic time (時), ritual significance (特), social hierarchies (侍), and artistic expression (诗). This progression from physical to abstract domains demonstrates the sophisticated metaphorical thinking in Chinese character formation.

## 5.2 Case Studies - Pinyin - A Counter-Argument

In the Chinese language, sound (声 shēng), form (形 xíng), and meaning (意 yì) are all integral components of a vibrant and living system. Over-emphasizing any single aspect at the expense of the others would be inefficient, misguided, and unwise. While the Pinyin romanization system has undeniably augmented the Chinese language by integrating Latin phonetic components, relying solely on Pinyin and abandoning Chinese characters would result in a tremendous loss of semantic meaning and cultural wisdom. The limitations of a purely phonetic system become apparent below using cases of homophony:

1. Single-character homophony:
   - mā: 妈 (mother), 蚂 (ant), 马 (mǎ, horse), 骂 (mà, yell or curse)
   - shì: 是 (to be), 事 (matter), 市 (market), 式 (style), 世 (world)
   - yì: 意 (meaning), 义 (righteousness), 艺 (art), 易 (change)
2. Compound-word disambiguation:
   - "xiansheng" in Pinyin could represent:
     - 先生 (teacher/mister)
     - 献身 (sacrifice oneself)

Chinese characters, in their elegance and complexity, encode both visual and auditory information while preserving crucial semantic distinctions. This harmonious integration of form and function has sustained the language for millennia. The path forward lies in preserving this sophisticated writing system while thoughtfully incorporating modern tools like Pinyin and AI as complementary aids rather than replacements.

## 5.3 Case Studies - Characters and Phrases

Just as matter organizes itself into increasingly complex structures - from atoms to molecules to molecular clusters - language exhibits similar emergent properties at different scales. Individual characters (字) serve as the atomic units, carrying fundamental meanings and combining properties. These form compounds and phrases (词组), analogous to molecules with stable semantic bonds. At a higher level of organization, these linguistic molecules arrange themselves into sophisticated structures like poems, which, like molecular clusters, exhibit properties beyond the sum of their parts. This natural hierarchy of meaning-making demonstrates the living and self-organizing nature of Chinese language.

### 5.3.1 The 子 Family

| Formula | Meaning | Natural Insight | Pinyin |
|---|---|---|---|
| 女 + 子 = 好 | good, well, fine | Child with mother represents fundamental goodness | hǎo |
| 耂 + 子 = 孝 | filial piety | Elder above child shows respect and care | xiào |
| 子 + 小 = 孙 | grandson | Small child represents generational continuation | sūn |
| 木 + 子 = 李 | plum tree | Fruit as nature's offspring | lǐ |
| 米 + 子 = 籽 | seed | Grain's offspring, agricultural reproduction | zǐ |
| 禾 + 子 = 季 | season | Crop cycle marked by growth stages | jì |
| 宀 + 子 = 字 | character | Child under roof represents nurturing of knowledge | zì |
| 小 + 冖 + 子 = 学 | to learn, to study | Child (子) under shelter (冖) starting small (小) captures the essence of education | xué |

The character 子 (zǐ), originally depicting a child with outstretched arms, functions as both a semantic and phonetic component in character formation. Its visual evolution preserves the core meaning of offspring while extending into broader domains of growth and nurturing.

### 5.3.2 The 子 Network

| Category | Compounds | Semantic Extension | Examples |
|---|---|---|---|
| Human Relations | 父子, 子女, 子孙, 弟子 | Core family ties | father-and-son, children, descendants, disciple |
| Honorific Terms | 老子, 孔子, 墨子, 孙子, 君子 | Respect and wisdom | Master, great teachers, gentlement |
| Scientific Objects | 量子, 光子, 原子, 电子 | Fundamental particles | Quantum, photon, atom, electron |
| Mathematical Objects | 因子, 子集, 子空间 | Math concepts | factor, subset, subspace |
| Natural Elements | 种子, 脑子, 芽子, 子宫 | Biological growth | Seed, brain, sprout, womb |
| Animals | 狮子, 兔子, 蚊子 | living objects | Lion, rabbit, mosquito |
| Tools/Objects | 筷子, 梯子, 桌子, 房子, 子弹 | Functional items | Chopsticks, ladder, table, house, bullet |
| Time Concepts | 日子, 子时, 甲子 | Temporal cycles | Days, midnight hour, 60-year cycle |
| Aggressor | 日本鬼子, 毛子 | Nicknames | Japanese soldier, northern invader |

This analysis reveals how 子 extends from its concrete meaning of "child" into an extraordinarily rich semantic network spanning natural, social, and abstract domains. At the foundational level, it captures core human relationships (父子, 子女) and biological growth (种子, 子宫). Its semantic scope then expands remarkably to encompass everything from the cosmic scale of fundamental particles (量子, 原子) to the precision of mathematical abstractions (因子, 子集). The character's versatility is further demonstrated in its application to everyday objects (筷子, 房子), temporal concepts (日子, 甲子), and even cultural-historical contexts (日本鬼子). What's particularly striking is how 子 maintains its core associations with growth, fundamentality, and belonging across these diverse domains - whether describing a physical descendant (孙子), an intellectual lineage (弟子), or the smallest units of matter (原子). This semantic radiation from concrete to abstract meanings, while preserving core conceptual links, exemplifies the sophisticated metaphorical thinking embedded in Chinese character evolution and compound formation.

## 5.4 Case Studies - Poems: Temple of Words

This section examines how classical Chinese poetry achieves remarkable semantic density through minimal character usage. The survival of these poems across millennia demonstrates powerful principles of cultural selection, where maximum meaning with minimum structure creates enduring linguistic artwork.

### 5.4.1 Poetry as Sacred Architecture

讠 + 寺 = 诗: This elegant formation captures the essence of poetry - the construction of linguistic temples where carefully chosen words create spaces of meaning that transcend their components. Just as temples transform physical space into sacred realm through architectural principles, classical Chinese poems create meaning through precise structural arrangement.

When this author queries AI systems for exemplary Chinese poems, these two masterpieces emerge with remarkable frequency, suggesting their special status in the corpus of Chinese literature.

(III) 王之涣 - 登鹳雀楼

白日依山尽，
黄河入海流。
欲穷千里目，
更上一层楼。

(IV) 李白 - 静夜思

床前明月光，
疑是地上霜。
举头望明月，
低头思故乡。

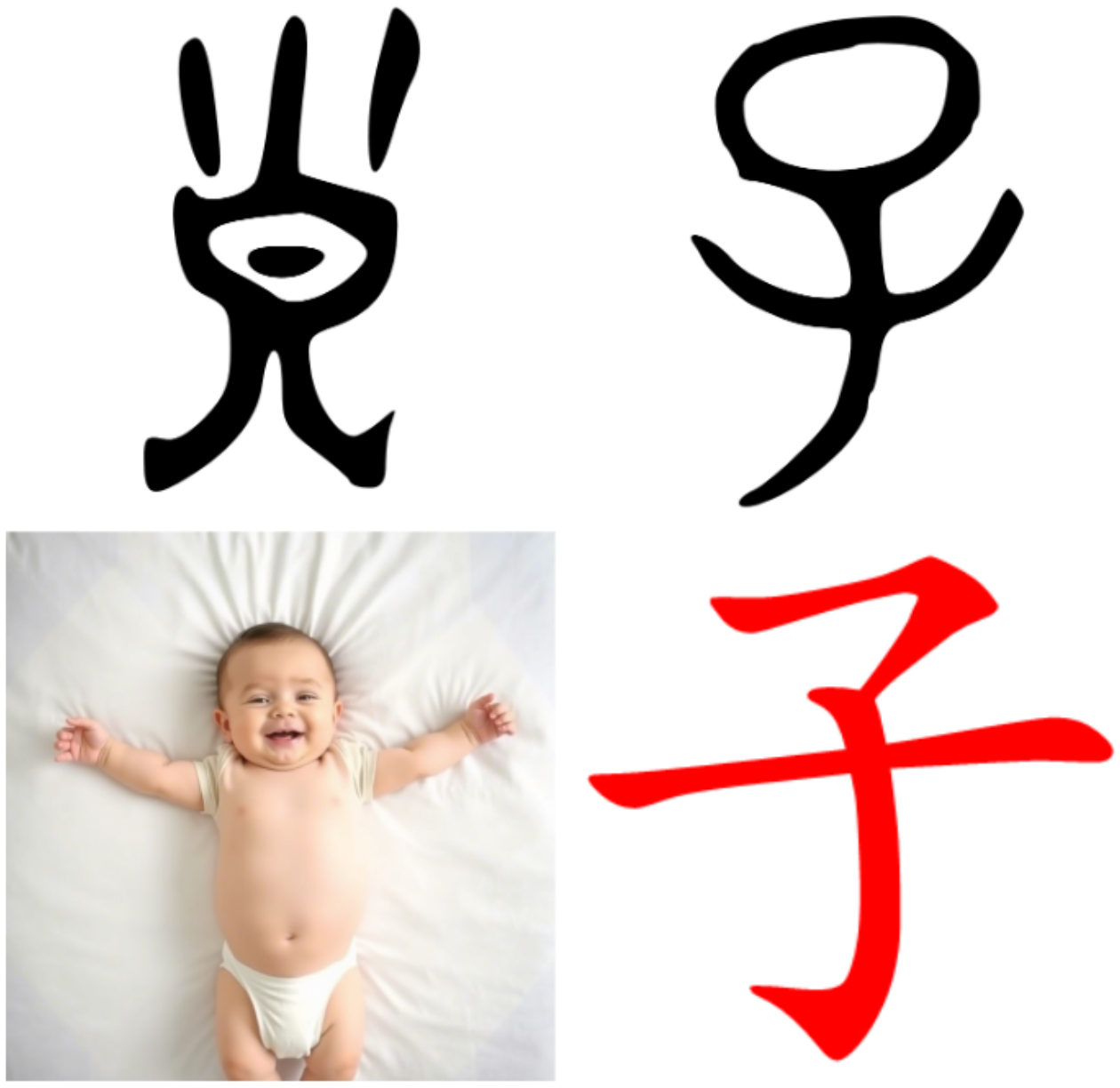

*Figure 12: Evolution of character 子 (child) from oracle bone script to modern form, with visual metaphor of a baby with outstretched arms.*

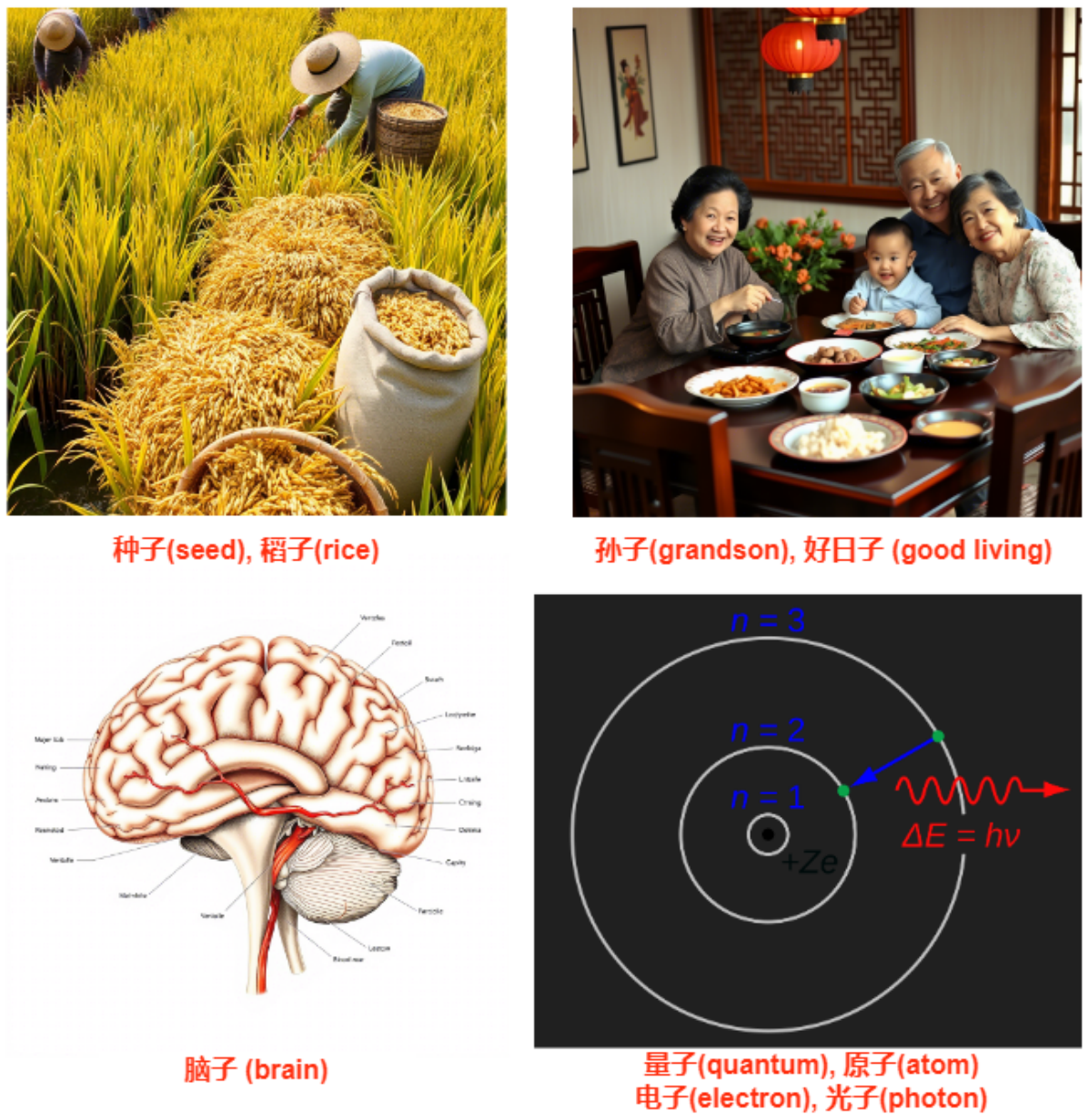

*Figure 13: Semantic extensions of 子 across various domains [12].*

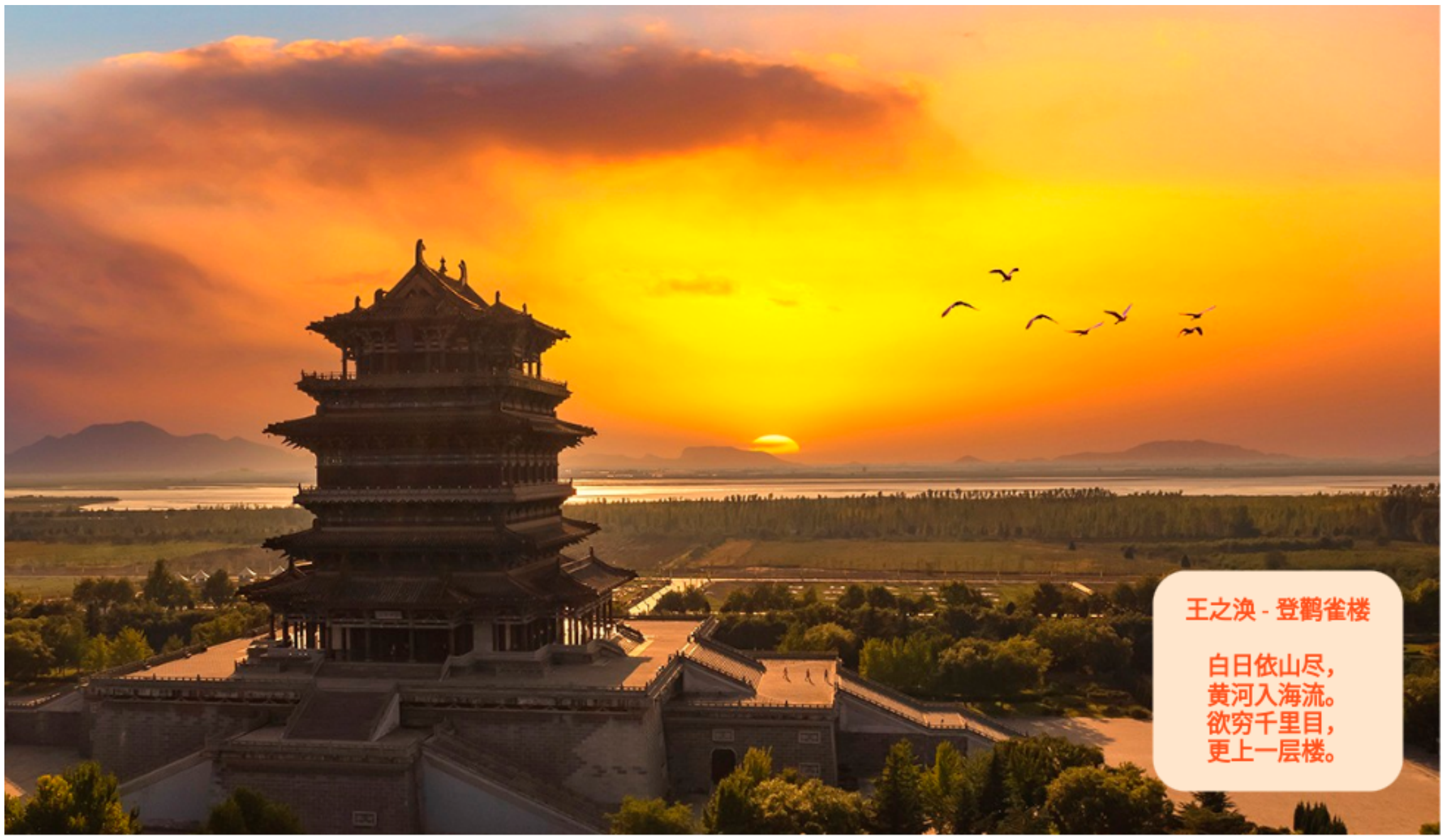

*Figure 14: A magnificiant view of Huang-He-Lou* 鹳雀楼.

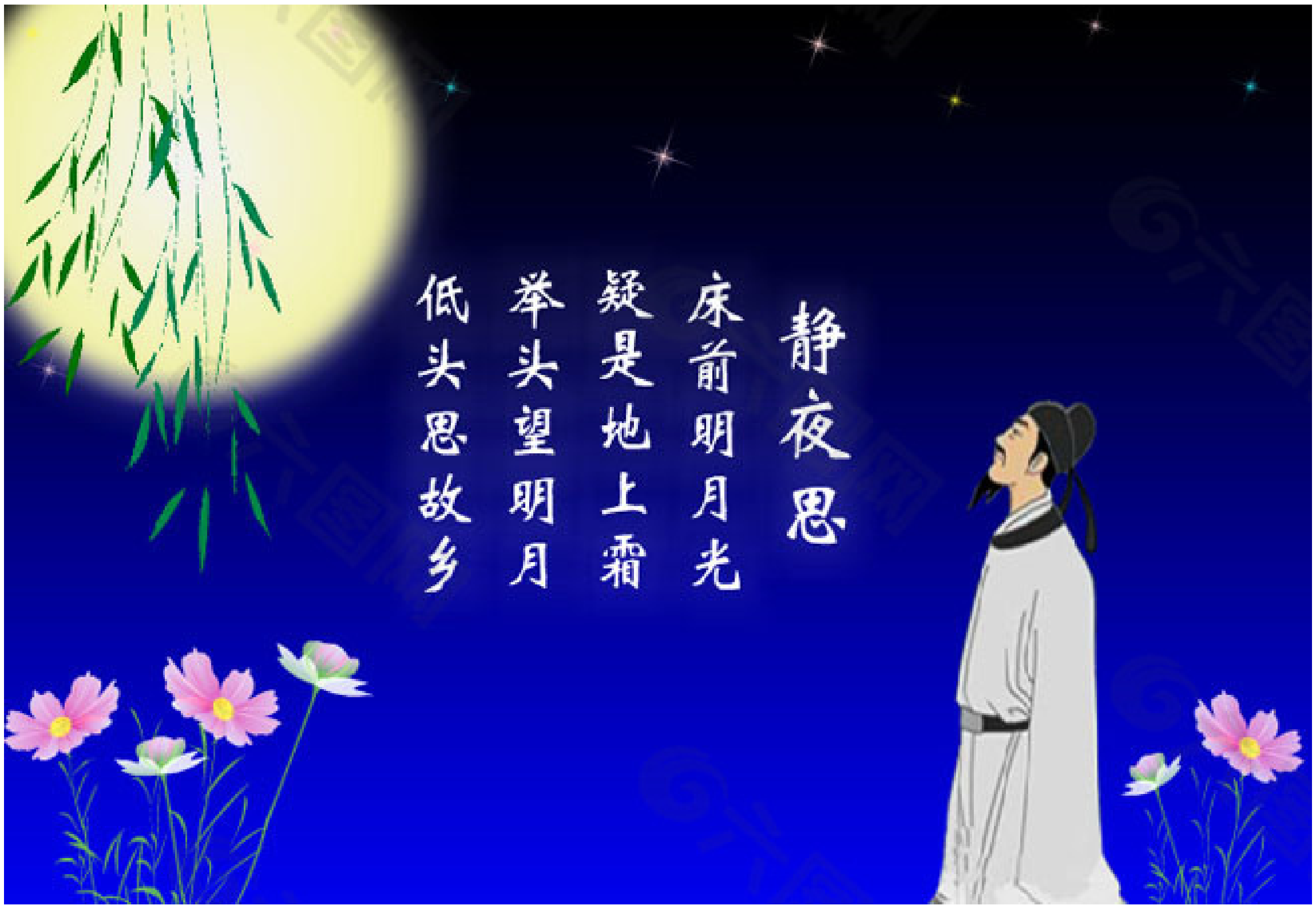

*Figure 15: An artistic illustration depicting Li Bai in deep comtemplation under moonlight.*

### 5.4.2 Architectural Principles of Poetic Construction

#### 5.4.2.1 Foundation (地基)

- Elementary Characters as Building Blocks
  - Natural elements: 山, 河, 日, 月
  - Basic actions: 上, 望, 举, 低
  - Core concepts: 目, 光, 头, 楼

#### 5.4.2.2 Vertical Progression (层进)

- Physical and Spiritual Elevation
  - "登鹳雀楼": Literal ascent mirrors expanding consciousness
  - "静夜思": Movement from earth (霜) to heaven (月) to heart (思)
- Each level adds new dimension of meaning

#### 5.4.2.3 Sacred Space (空间)

- Both poems create vast mental landscapes through minimal means:
  - Horizontal expanse: 千里目, 入海流
  - Vertical dimension: 一层楼, 举头望
  - Internal realm: 思故乡

### 5.4.3 Yin-Yang Duality in Expression

These poems together form a complete cognitive framework through complementary approaches:

| **Aspect** | **Yang (阳) - 登鹳雀楼** | **Yin (阴) - 静夜思** |
|---|---|---|
| Celestial Bodies | Sun setting (白日依山) | Moon reflecting (明月光) |
| Movement | Upward progression (更上一层楼) | Circular motion (举头...低头) |
| Philosophy | Active pursuit of transcendence | Passive reception of insight |
| Emotion | Aspiration toward horizons | Nostalgia and connection |

### 5.4.4 Evolutionary Fitness Factors

The remarkable survival and transmission of these poems can be attributed to several adaptive advantages:

##### 5.4.4.1 Structural Efficiency

- Semantic Density: Maximum meaning in minimum space (20 characters)
- Progressive Construction: From observation to insight
- Information Compression: Multiple layers of meaning per character
- Mnemonic Design: Rhythm and imagery supporting memory

##### 5.4.4.2 Cultural Transmission Vectors

- Educational Value
  - Accessible characters (易懂) with sophisticated craft (难工)
  - Clear structure supporting memorization (朗朗上口)
  - Universal themes ensuring relevance (代代相传)
- Cognitive Resonance
  - Alignment with fundamental thought patterns
  - Balance of concrete and abstract elements
  - Integration of emotion and insight

**5.4.4.3 Complementary Wisdom Paths** These masterpieces encode two essential approaches to understanding:

- "登鹳雀楼": Transcendence Through Effort
  - Physical elevation as metaphor for understanding
  - Active engagement with limitations
  - Expansion of perspective through deliberate action
- "静夜思": Insight Through Reflection
  - Quiet observation revealing deep connection
  - Natural phenomena evoking emotional truth
  - Stillness as path to profound understanding

#### 5.4.5 Summary Remark

These poems demonstrate the sophisticated architecture of classical Chinese poetry, where careful arrangement of minimal elements creates enduring structures of meaning. Their survival through millennia reflects both their remarkable efficiency in encoding wisdom and their resonance with fundamental patterns of human experience and understanding. Like well-designed temples, they transform simple elements into spaces of deep meaning, achieving maximum impact on culture and civilization.

# 6 Natural Evolution of Characters

In our earlier sections on “Visualizing Categorization” and “Character Compositional Patterns”, we observed how elemental characters reflect human cognition about nature, how elemental characters interact like atomic objects in the physical world. Here we explore how Chinese characters may have evolved naturally.

## 6.1 Historical Context

According to legend, Cangjie (仓颉) invented Chinese characters, who was a historian to the Yellow Emperor of China in the 27th century BCE.

Cangjie was inspired by the natural world, including animal tracks, landscapes, and the stars. He believed that if he could capture the unique features of everything in a single painting, he could create a writing system. He was frustrated with the limitations of knotting, which was the previous method of recording information.

As a milestone event, Cangjie undoubtedly contributed to the creation of Chinese writing system. In the grand scheme of a human language, Chinese characters may have evolved naturally, generation after generation, contratory to popular belief that it was a magic artificial design by Cangjie.

## 6.2 Natural Growth Patterns in Character Evolution

Just as natural systems evolve from simple to complex through predictable patterns, Chinese characters demonstrate similar organic development. The Fibonacci sequence (1,1,2,3,5,8,13,21,34,...) provides an elegant framework for understanding this evolution, not as a rigid mathematical correspondence, but as a metaphor for how complexity emerges from simplicity in systematic ways.

The accompanying images of Fibonacci patterns in nature - from sunflower seed arrangements to nautilus shells, from fern fronds to spiral galaxies - reveal this universal principle of growth and organization. This same principle can illuminate our understanding of how Chinese characters evolved from Cangjie’s initial inspiration to their current form.

Inspired by universal natural growth pattern like Fibonacci sequence, this author borrows the sequence to organize Chinese characters in a similar natural progression: from simple pictographs to sophisticated composite characters, from concrete objects to abstract concepts. This natural progression is not just an organizational framework - it reflects how human cognition itself evolved from basic perceptions to complex abstractions. Just as the Fibonacci spiral demonstrates how complex natural patterns emerge from simple mathematical relationships, our organization reveals how Chinese writing evolved from basic elements (元字) into increasingly complex expressions. Each level introduces new fundamental characters that, like the expanding spiral, serve as building blocks for richer linguistic representations. This approach mirrors nature’s own efficiency in developing complex systems from simple foundations.

## 6.3 Elemental Character (元字) Levels

In the following, we list the 元字 collection from first 9 levels Traditional radicals with independent meaning are noted under “Radical form mapping” (e.g. 灬 means fire).

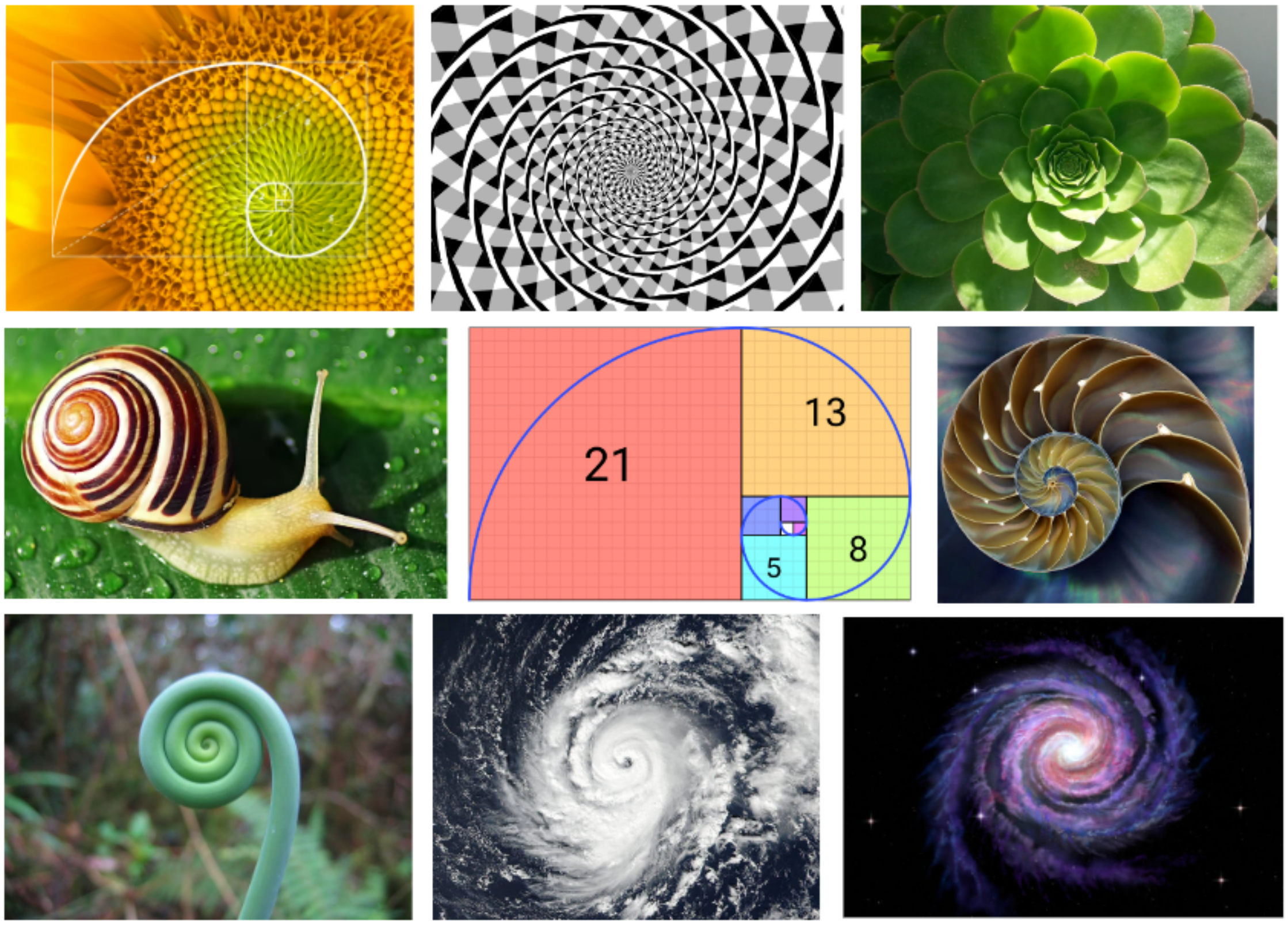

*Figure 16: Natural manifestations of the Fibonacci spiral across scales: botanical (sunflower, succulent), biological (snail shell, fern), mathematical (golden ratio), and astronomical (galaxy) examples.*

#### 6.3.1 1 (一): 气 (primordial force/energy)

- The most fundamental 元字
- Represents the emergence of invisible form from formlessness (无中生有)
- Base unit for energy and force concepts
- 炁 is an uncommon and old form for 气, rarely used. But its lower radical (灬) hints its semantic meaning related to fire and energy.
- Radical form mapping: 气 is a radical, all characters containing 气 are related to fundamental nuclear and chemical elements or matter in gaseous form, e.g., 氢 (hydrogen gas), 氦 (Hellium), 氧 (oxygen), 汽 (vapor), 氛 (atmosphere)

#### 6.3.2 1 (一): 点，线 (primitive low-dimensional object)

- Basic radicals 元字: 丶 (dot), 一 (horizontal line), 丨 (vertical line), 丿 (north-east line), ㇏ (south-east line)
- Represents visible simple form (i.e., point- or line-like objects), although carrying no indenpendent semantic meaning, they derive meaning together with other composing character part. As an example, we build upon the first elemental character 气,
  - 气 + 丿 = 氕 is Protium, the most common and stable isotope of hydrogen with one proton (i.e., 1 nucleon);
  - 气 + 刂 = 氘 is Deuterium, stable and 0.02% of naturally occurring hydrogen, with one proton and one neutron (i.e., 2 nucleons);
  - 气 + 川 = 氚 is Tritium, a radioactive isotope of hydrogen, with one proton and two neutrons (i.e., 3 nucleons). Here, 丿, 刂, 川 indicate 1,2,3 nucleons in the respective hydrogen isotopes.

#### 6.3.3 2 (二): 日, 月 (sun and moon)

- First pair of naturally contrasting 元字
- Represents 2 visible solar objects and a fundamental abstraction in the basic dualism philosophy (阴阳)
- Foundation for temporal and luminance concepts
- Both characters can be used as radicals. It is worthwhile to note that 月 means body part meat/flesh 肉 when used as radical. This is likely a historical coincidence where 月 was adopted as the simplified writing form for 肉.

#### 6.3.4 3 (三): 天, 地, 人 (heaven, earth, human)

- Tripartite domain 元字
- Establishes basic spatial and existential framework for human cognitive psychic
- Core reference for positioning and relationships
- Radical form mapping: 土 for 地, 亻 for 人.

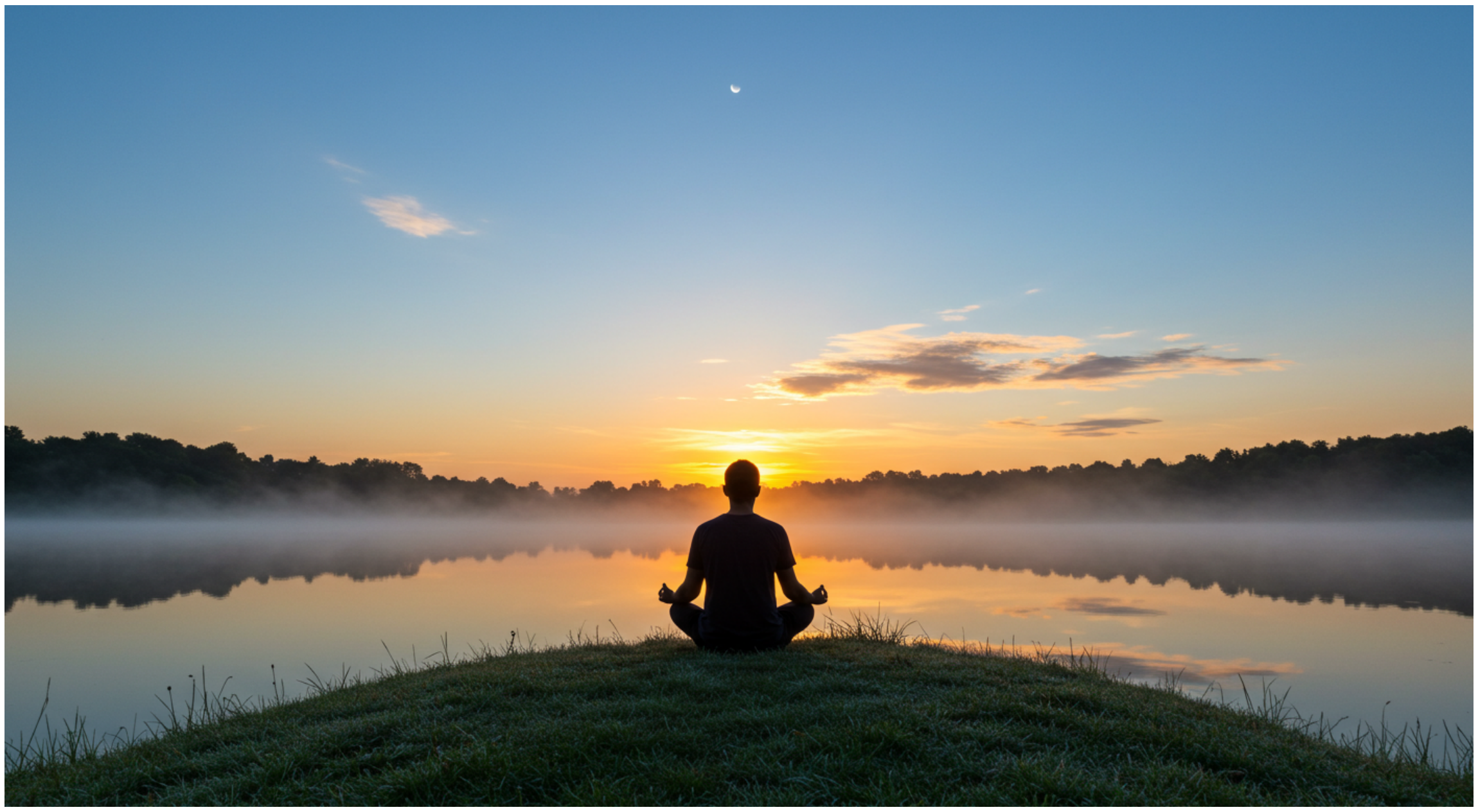

*Figure 17: AI-generated composition depicting the harmony between human consciousness and nature, incorporating six elemental characters (气, 日, 月, 天, 地, 人).*

#### 6.3.5 5 (五): 金, 木, 水, 火, 土 (metal, wood, water, fire, earth)

- Material phase 元字
- Fundamental 5 elements (五行) for describing physical and materialistic world in ancient philosophy.
- Base components for nature-related characters
- Radical form mapping: 钅 for 金, 氵冫 for 水, 灬 for 火, 木, 土 are often rended in narrower form when used as radicals. The semantic meanings are the same.

#### 6.3.6 8 (八): 东, 南, 西, 北, 春, 夏, 秋, 冬 (directions and seasons)

- Spatiotemporal 元字
- Complete system of orientation and cyclical change
- Foundation for location/directional and time-based concepts

#### 6.3.7 13 (十三): 生, 鼠, 牛, 虎, 兔, 龙, 蛇, 马, 羊, 猴, 鸡, 狗, 猪 (basic life forms expressed in 12 Zodiac animals)

- Biological object 元字
- Complex natural phenomena
- Base set for describing living things
- Radical form mapping: 牜 for 牛, 虫 is a radical for 蛇 and other insects, 犭 is a radical for many animals (e.g. 猴, 狗, 猪), 羊, ⺶, ⺷ are variant radical forms for 羊 (Sheep), radical for 马 appears narrower.

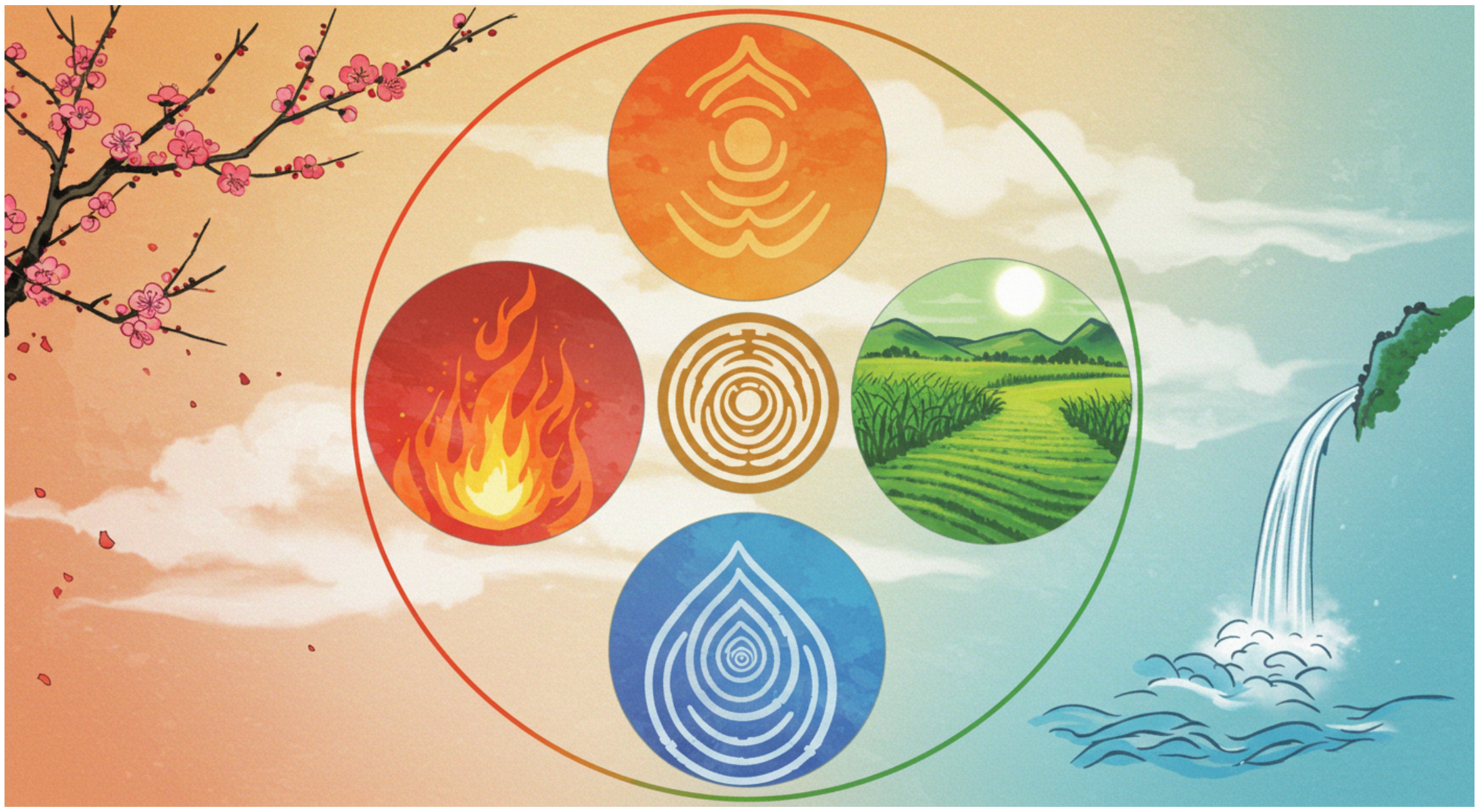

*Figure 18: This AI generated [9] artistic representation of the Five Elements (五行) in Chinese philosophy, emphasizing the dynamic and cyclical nature of these interconnected elements in early cosmology.*

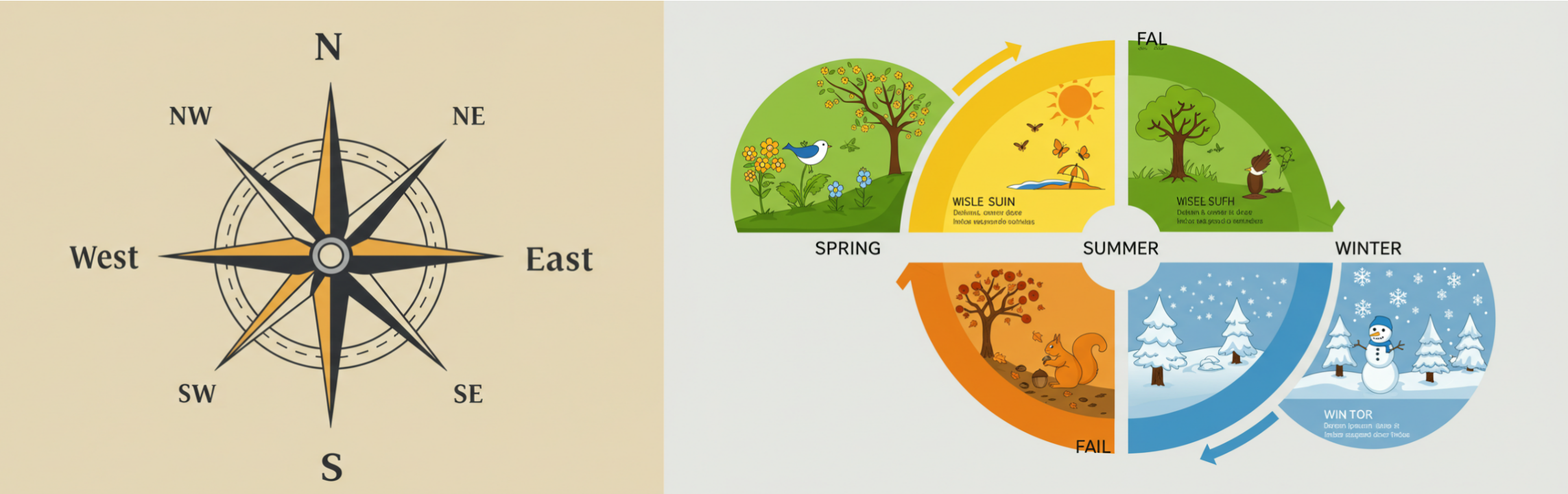

*Figure 19: This AI generated [9] view of fundamental spatial and temporal orientations. Left: A traditional compass rose showing the four cardinal directions (North, South, East, West) and four intercardinal directions (NE, SE, SW, NW); Right: A circular diagram depicting the cyclical nature of the four seasons.*

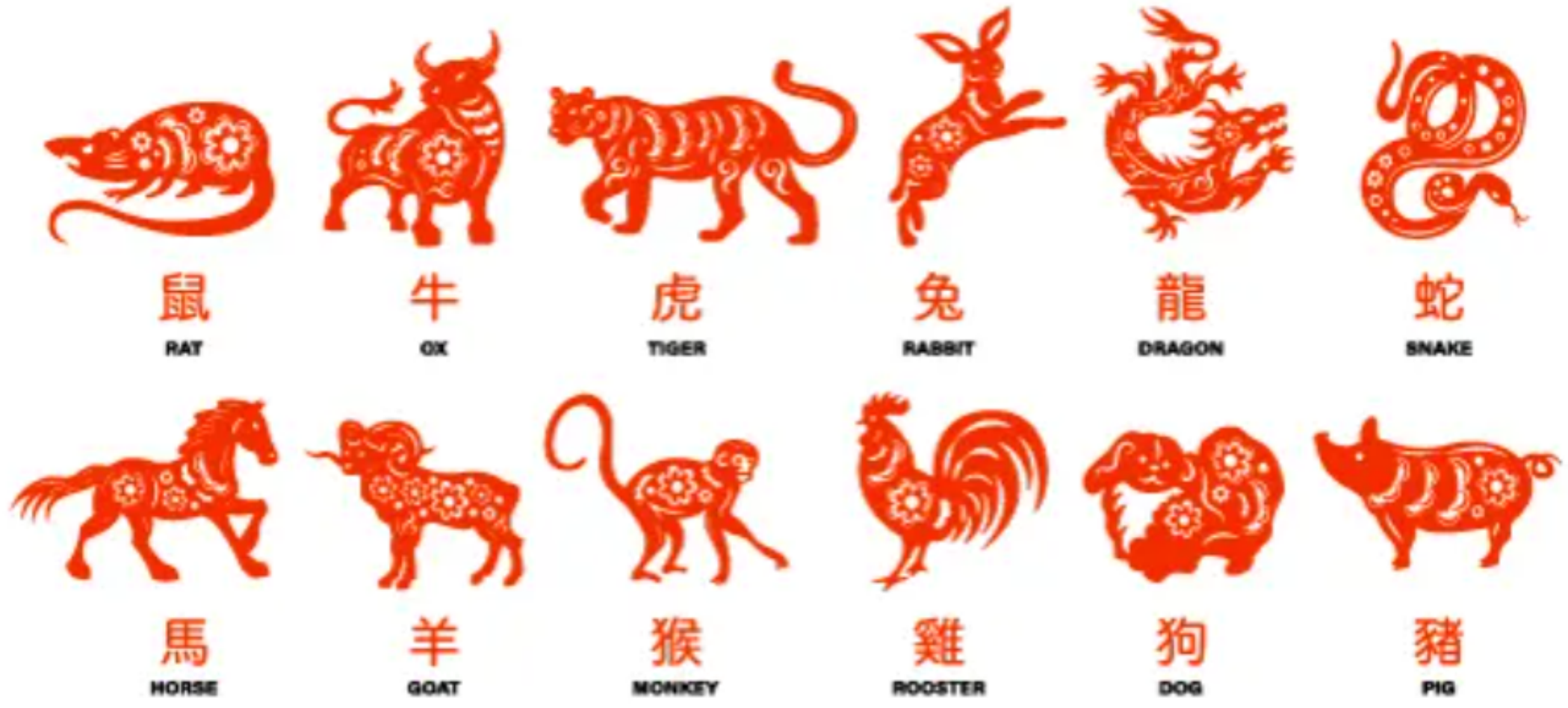

*Figure 20: These twelve animals represent the complete cycle of the Chinese zodiac (十二生肖), with each animal corresponding to a year in the twelve-year astronomical cycle.*

#### 6.3.8 21 (二十一): Quantification and Measurement 元字

- Numerical System (15 characters):
  - Basic numerals: 一 (1), 二 (2), 三 (3), 四 (4), 五 (5), 六 (6), 七 (7), 八 (8), 九 (9), 十 (10)
  - Large quantities: 百 (100), 千 (1000), 万 (10,000), 亿 (100,000,000), 零 (0)
  - These form the foundation for all quantitative description
- Physical Units (6 characters):
  - Time measurement: 秒 (s), 分 (m), 时 (h)
    * Progression from smallest (second) to largest (hour)
    * Reflects natural cycles and human activity patterns
  - Length measurement: 寸 (cm), 丈 (m), 里 (km)
    * Traditional Chinese units of length
    * Scales from human body reference (寸) to geographic distance (里)

This level represents the emergence of systematic measurement and counting.

#### 6.3.9 34 (三十四): Human Form and Action 元字

- Basic parts: 心 (忄), 头, 首, 面, 口, 目, 眉, 鼻, 耳, 舌, 牙, 齿, 手 (扌), 又, 足, 血, 肉, 身, 尸, 骨, 皮, 毛 (彡)
- Action indicators: 言 (讠), 看, 听, 思, 食 (饣), 走 (辶), 立
- Identity: 男, 女, 子, 自, 己
- Radical form mapping: 忄 for 心, 扌, 又 for 手, 辶 for 足, 讠 for 言, 饣 for 食, often (not always) in action context, e.g. emotional thinking, holding, walking, communicating, respectively. 目, 口, 足, 骨, 耳 appear as radicals too. Some of these characters (like 首, 面) can function as both nouns (head, face) and measure words/classifiers in different contexts.

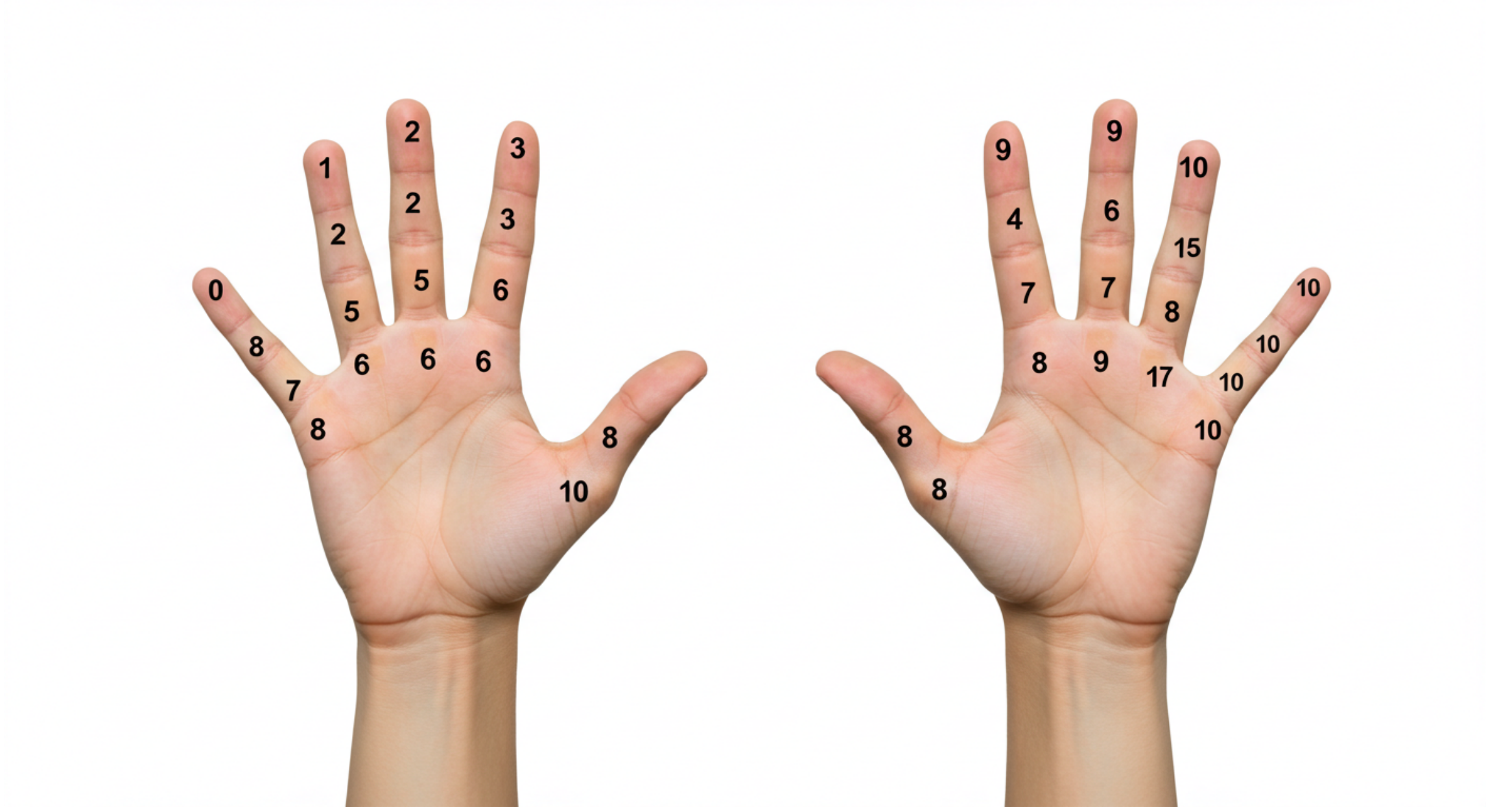

*Figure 21: The anatomical basis of counting may explain why base-10 systems emerged independently across various civilizations and why the Arabic numeral system (0-9) became globally dominant over other numerical systems in history.*

This level introduces fundamental components for describing human existence and behavior.

In summary, these nine levels demonstrate how Chinese characters evolved naturally from fundamental concepts to sophisticated human expression, mirroring the complexity gradients found in natural systems.

## 6.4 Discussion and Implications

### 6.4.1 Chinese Writing as a Living System

Our computational and physics-inspired analysis reveals Chinese writing as a living, self-organizing system that mirrors natural growth patterns. This perspective yields several key insights:

#### 6.4.1.1 Emergent Complexity

- Like biological systems emerging from simple molecular interactions, complex characters emerge from basic 元字 combinations
- Component relationships evolve naturally based on semantic needs, not arbitrary assignments
- New meanings emerge through systematic combination patterns
- Character evolution follows principles similar to natural growth patterns seen in Fibonacci sequences
- The progression from basic elements (气, 点, 线) to sophisticated human concepts demonstrates organic development

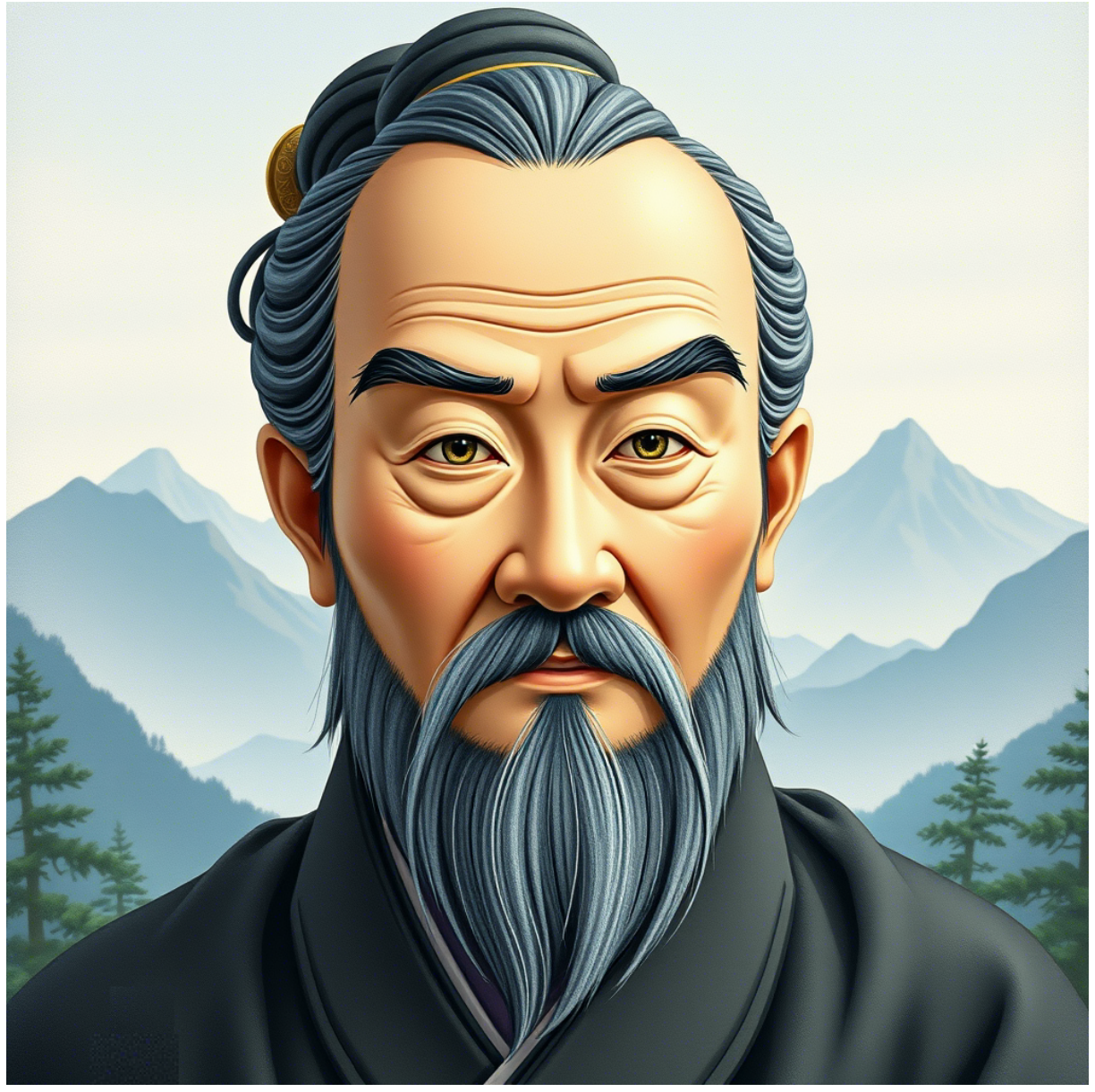

*Figure 22: An AI-generated [12] artistic portrait depicting Laozi (老子, also known as Lao Tzu), the legendary 6th century BCE Chinese philosopher and reputed author of the Tao Te Ching (道德經).*

#### 6.4.1.2 Self-Organization Principles

- Characters form stable patterns without centralized design
- Component combinations follow natural efficiency principles
- Semantic relationships develop through organic usage patterns
- System exhibits both stability (preserving core meanings) and adaptability (generating new combinations)
- Character families demonstrate natural clustering and relationship patterns

#### 6.4.1.3 Adaptive Resilience

- System maintains coherence while allowing innovation
- Character combinations show remarkable flexibility in expressing new concepts
- Ancient elements remain relevant for modern technical terms
- Core semantic foundations support endless expansion

### 6.4.2 Practical Implications

#### 6.4.2.1 Learning Strategy

- Focus on elemental character (元字) and their core concepts
- Reduce the fundamental 元字 set to lower the entry-barrier in learning Chinese
- Learn the set of 400 元字 first as foundation

- Recognize systematic structure instead of memorizing thousands of individual characters
- Understand natural combination patterns rather than arbitrary associations

#### 6.4.2.2 Educational Applications

- Promote concept-based learning across disciplines
- Break down artificial barriers between language, mathematics, and science
- Make STEM-oriented learning accessible to earlier ages through character analysis
- Use character evolution to teach systematic thinking
- Leverage natural patterns to enhance memorization and understanding

#### 6.4.2.3 Modern Applications

- Character design principles for digital fonts
- Input method optimization based on component patterns
- Natural language processing models based on character component relationships
- AI applications in character recognition and generation
- Development of new technical terminology
- Cross-cultural communication of scientific concepts

### 6.4.3 Basic Research in Other Natural Languages

The methodologies and insights developed in this analysis of Chinese characters suggest promising directions for studying other natural languages.

#### 6.4.3.1 Computational Approaches

- Apply network analysis to study semantic relationships and word formation patterns
- Investigate natural clustering and organizational principles in vocabulary
- Analyze language evolution through computational models
- Map concept hierarchies across different languages

#### 6.4.3.2 Physics-Inspired Analysis

- Use reductionist approaches to identify fundamental linguistic elements
- Study language as a complex adaptive system
- Apply principles of self-organization to understand language evolution
- Investigate universal patterns in language structure

#### 6.4.3.3 Enhanced Learning Frameworks

- Simplify language learning by identifying core patterns and elements
- Create concept-based learning approaches that transcend specific languages
- Develop multi-modal learning experiences that leverage natural associations
- Build cross-linguistic bridges through shared conceptual foundations

#### 6.4.3.4 AI-Enhanced Applications

- Leverage AI for personalized, multi-lingual learning experiences
- Create interactive visualizations of language relationships
- Develop tools for concept-based cross-language translation
- Enable dynamic, context-aware language learning environments

This analysis suggests that viewing languages as living, evolving systems rather than fixed sets of rules opens new possibilities for learning, teaching, and research. The natural principles revealed in Chinese character formation and evolution can inform both theoretical understanding and practical applications across multiple languages and disciplines.

# 7 Conclusion

This research demonstrates how simplification and deeper understanding can work hand in hand in the study of Chinese characters. By identifying 422 elemental characters (元字) through computational network analysis, we have achieved significant simplification of the learning challenge - reducing the initial memorization burden from thousands of characters to a manageable set of foundational elements. However, this simplification leads not to reduction but to enrichment of understanding in several key ways:

(1) From Memorization to Understanding

- Instead of rote learning of thousands of isolated characters
- Learners grasp systematic patterns of character formation
- Understanding how complex meanings emerge from simple elements
- Recognition of natural organizational principles in language

(2) From Static to Dynamic Understanding

- Characters revealed as a living, evolving system
- Clear visualization of how meanings extend from concrete to abstract
- Appreciation of the system's adaptive capacity for new concepts
- Recognition of internal logic in character evolution

(3) From Fragmented to Integrated Knowledge

- Connection between language and natural patterns
- Integration of linguistic, philosophical, and scientific insights
- Bridge between traditional wisdom and modern analysis
- Cross-disciplinary understanding of knowledge organization

(4) From Surface to Deep Structure

- Beyond simple visual components to semantic networks
- Recognition of sophisticated metaphorical thinking
- Understanding of systematic meaning extension
- Appreciation of cultural and cognitive patterns

This deeper understanding enriches learning experiences by:

- Making character learning more intuitive and systematic
- Revealing connections across different knowledge domains
- Enabling appreciation of cultural and philosophical dimensions
- Supporting creative engagement with the writing system

The research demonstrates that true simplification comes not from reduction alone, but from revealing underlying patterns and principles. By understanding Chinese characters as a naturally evolved system, we gain both practical advantages in learning and deeper insights into how human cognition organizes and transmits knowledge.

This approach opens new possibilities for:

- AI-enhanced learning tools based on natural patterns
- Cross-cultural communication of complex concepts
- Integration of traditional wisdom with modern analysis
- Development of more intuitive teaching methods

The success of this analysis suggests that the path from simplification to deeper understanding is not a trade-off but a synergy, where reduced complexity reveals deeper patterns and richer meanings. This insight has implications not only for Chinese language learning but for how we approach the study of complex systems in general.

## 8 Dedication

This work is dedicated to late Professor T.D. Lee（李政道）, whose pioneering efforts opened doors for many Chinese students to pursue studies and research in the United States, fostering a bridge between Eastern and Western scientific traditions. His vision and support have enabled countless scholars like myself to contribute to global scientific discourse.

This work is also dedicated to author's parents (龚永权，文国芳), who nurtured his intellectual curiosity through hardships and challenging times. Their sacrifices and unwavering support are forever remembered.

## 9 Acknowledgements

This paper represents a collaborative effort between the author, and a team of AI assistants including Claude 3.3 from Anthropic [10], Gemini 2 from Google [11], Qwen2.5-Max from Alibaba [12]. The fusion of human knowledge and insight in software development, physics, and Chinese language with AI's analytical capabilities enabled the development of the novel perspectives and methodologies presented in this work. This collaboration demonstrates the potential of human-AI partnerships in research and learning, particularly in interdisciplinary studies bridging traditional knowledge with modern computational technologies.